\documentclass{llncs}

\makeatletter
\newif\if@restonecol
\makeatother

\usepackage[linesnumbered,ruled,algonl,vlined]{algorithm2e}

\usepackage{graphicx}
\usepackage{epstopdf}
\DeclareGraphicsExtensions{.pdf,.eps,.png,.jpg,.mps}
\usepackage{placeins} 

\newcommand{\eat}[1]{}

\begin{document}


\title{Shared Execution of Path
Queries on Road Networks}

\author{
Hossain Mahmud
\and Ashfaq Mahmood Amin
\and Mohammed Eunus Ali
\and Tanzima Hashem
}

\institute{Department of Computer Science and Engineering\\ Bangladesh University of Engineering and Technology, Dhaka 1000, Bangladesh\\
\email{\{hossain.mahmud, ashfaq.m.amin\}@gmail.com, \{eunus, tanzimahashem\}@cse.buet.ac.bd } }

\date{October 28, 2012}

\maketitle
\begin{abstract}
The advancement of mobile technologies and the proliferation of map-based applications  have enabled a
user to access a wide variety of services that range from information queries to navigation systems. Due to the popularity of map-based applications among the users, the service provider often requires to answer a large number of simultaneous queries. Thus, processing queries efficiently on spatial networks (i.e., road networks) have become an important research area in recent years. In this paper, we focus on path queries that find the shortest path between a source and a destination of the user. In particular, we address the problem of finding the shortest paths for a large number of simultaneous path queries in road networks. Traditional systems that consider one query at a time are not suitable for many applications due to high computational and service costs. These systems cannot guarantee required response time in high load conditions. We propose an efficient group based approach that provides a practical solution with reduced cost. The key concept for our approach is to group queries that share a common travel path and then compute the shortest path for the group. Experimental results show that our approach is on an average ten times faster than the traditional approach in return of sacrificing the accuracy by 0.5\% in the worst case, which is acceptable for most of the users.
\end{abstract}

\keywords{Spatial database, Query processing, Road networks, Clustering}

\section{Introduction}

With the proliferation of GPS-enabled mobile technologies, users access a wide variety of location-based services (LBS) from different service
providers. These LBS range from simple information queries such as finding the nearest restaurant to navigational queries such as finding the
shortest path to a destination. In this paper, we focus on path queries that find the shortest path between a source and a destination of the user.
In particular, we address the problem of finding the shortest paths for a large number of simultaneous path queries in road networks. Traditional
systems that consider one query at a time are not suitable for many applications as these systems cannot guarantee a cost-effective and real-time
response to the user in high load conditions~\cite{Levandoski:2010,Zhang:SmashQ}.  We propose an efficient group based approach that provides a
practical solution for path queries with reduced computational cost.

In a road network, users are often interested in a service (e.g., finding the restaurant or the path to the destination) that can be reached in
minimum travel time. Since travel time on a road segment is highly dynamic and depends on various real-time traffic
conditions~\cite{Demiryurek:2010}, it is not possible to accurately compute the travel time based on the network distance. Thus, to answer such user
queries, an LBS provider needs to gather real time traffic conditions of the underlying road networks.  However, it may not be possible for every LBS
provider to have their own monitoring infrastructure for real traffic updates. Therefore, to process queries on road networks, LBS providers
subscribe to map based services such as Google Maps~\cite{googlemaps}, MapQuest~\cite{mapquest}, Yahoo Maps~\cite{yahoomaps}, and Microsoft Bing
Maps~\cite{bingmaps} for traffic updates.

Due to huge client bases and popularity of map-based services, the LBS server often require to respond to a large number of \emph {simultaneous} user
queries. Thus, efficient processing of a large number of queries in road networks have become an important research area in recent years. Specially,
when an LBS server needs to call the map based services for every user request, it becomes a major bottleneck in providing a cost-effective and
real-time response to the user~\cite{Zhang:SmashQ}. The underlying reason of the problem is as follows. First, map based web services charge on per
request basis (e.g., in Google Maps~\cite{googlemaps}, an evaluation user can submit 2,500 requests per day and a licensed business user can submit
100,000 requests per day~\cite{googlemapapi}) and thus need to pay more for more user requests. Second, each map based web service call incurs a huge
delay in response, e.g.,  an web service call to fetch the travel time from the Microsoft MapPoint web service to a database engine takes 502
ms~\cite{Levandoski:2010}. In addition to the above problems,  well known approaches (e.g.,  Dijkstra~\cite{dijkstra} and A*
Algorithm~\cite{Nilson:Astar} ) for the shortest path computation require expensive graph traversal operations, and thus incur huge computational
overhead specially for a large number of queries. To alleviate all these problems, we propose a grouping approach for an LBS server that efficiently
process a large number of simultaneous path queries in road networks.

The key concept of shared execution of a group of path queries comes from the path coherence properties of  road networks~\cite{pathoracles}. Path
coherence is the concept that shows that the shortest paths from nearby sources to nearby destinations share a large common path among them. That is,
two spatially close source vertices $s_1;s_2$ share large common road segments of their shortest paths to reach two spatially close destination
vertices $d_1;d_2$, which are far from  $s_1;s_2$.  For example, it has been observed that the 30,000 shortest paths between two subsets of sources
and destinations in a road network of Silver Spring, MD pass through a single common vertex~\cite{pathoracles}.

Based on the above observation, to find a subgroup from a large number of path queries, we first find a cluster of source-destination pairs based on
similarities of the Q-lines, where a Q-line is defined as the connecting straight line between a source and destination of a given path query. We
introduce the \emph{distance function} to measure the similarities among Q-lines and the \emph{areas of influence} to prune the search space while
clustering similar Q-lines. These concepts form the bases of our clustering algorithm that returns a set source-destination regions' clusters. Each
cluster essentially is a group of path queries who have high probabilities of sharing a common travel path in their answers. After computing the
clusters, for each cluster, we compute the shortest path from a source region to its corresponding destination region, by only considering the
outgoing edges and incoming edges of the source and the destination regions, respectively. Finally, each individual path query is evaluated by
concatenating the three shortest paths: the shortest path from the source location to the starting point of the group shortest path, the group's
shortest path, and the shortest path from the end point of the group shortest path to the destination location.

Our group based heuristic to answer the shortest paths for a large number of simultaneous path queries significantly reduces the computational
overhead and ensure real-time response to the users. Though, the group based approach does not guarantee optimal shortest path for all queries, the
deviation from the optimal paths is found negligible. Extensive experimental study in a real road network show that our group based heuristic
approach is on an average ten times faster than the straightforward approach that evaluates each path query independently, in return of sacrificing
the accuracy by 0.5\%, which is acceptable for most of the users.

In summary, the contributions of this paper are as follows:
\begin{itemize}
    \item We formulate the problem of group based path queries in road networks.

    \item We develop an efficient clustering technique to group path queries based on similarities of Q-lines that form the base of our efficient solution to process a large number simultaneous path queries in a road network.
    \item We conduct an extensive experimental study to show the efficiency and the effectiveness of
    our approach.
\end{itemize}

\section{Related Works}

To handle a large number of queries in modern database systems, shared execution of queries have recently received a lot of
attention~\cite{AliTZK10,shareddb,sigmod12,dapd12}. The core idea of all these approaches is to group similar queries (i.e., who share some common
execution path) and then execute the group as a single query in the system. These approaches are found to be effective for many applications in
handling high load conditions, whereas traditional systems that consider one query at a time fail to deliver the required performance for such
applications. In this paper, we propose a shared execution approach for path queries on road networks, which is the first attempt of this kind.

The problem of finding the shortest path from a source to a destination on a graph (or spatial/road network) has been extensively studied in
literature (e.g.,~\cite{goldberg:silverstein,Nilson:Astar,dijkstra}). Dijkstra's~\cite{dijkstra} algorithm is the most well-known approach for
computing a single source shortest path with non-negative edge cost. Dijkstra~\cite{dijkstra} incrementally expands the search space, starting from
the source, along network edges until the destination is reached. Hence, Dijkstra's algorithm requires to visit nodes and edges that are far away
from the actual destination. To improve the efficiency of the above algorithm, few variations of Dijkstra's algorithm have been developed
(e.g.,~\cite{goldberg:silverstein}). An alternative school of thought employs hill climbing algorithms, A* search~\cite{Nilson:Astar} and
RBFS~\cite{book:norvigRBFS}, that use heuristics, e.g., Euclidian distances, to prune the search space. However, both Dijkstra's and hill climbing
algorithms incur expensive graph traversal operations and require complete recomputation on every update. These
approaches~\cite{goldberg:silverstein,Nilson:Astar,dijkstra} assume that the road conditions (e.g., traffic jam) will remain static.

To address the dynamic load conditions of road networks, several approaches have been proposed~\cite{lifelongPlanning,dynAstar,ramalingam:dsss}.
Dynamic SWSF-FP~\cite{ramalingam:dsss} does not regenerate new path for every update rather it only reconstructs the area affected by the update or
changes in the environment. However, the time complexity of this algorithm can be very high as it is proportional to the number of affected nodes.
Dynamic variants of A* search, such as, {\it dynamic anytime A*}~\cite{dynAstar} and {\it life long planning A*}~\cite{lifelongPlanning}, have also
been proposed to handle dynamic load conditions. The main idea behind these approaches is to  keep registered routes to overcome the problem of
complete recomputation. To compute the shortest path for dynamic edge costs, an algorithm called, dynamic single source shortest path (DSSS) is
proposed in~\cite{ramalingam:dsss}. This approach requires pre-computation of shortest paths for each source. King's approach for dynamic all pair
shortest path, APSP~\cite{king:fulldyn} requires pre-computation for each pair of nodes. The former one requires high memory space where as the
latter has the limitation that all edge weights need to be integers bounded by a small constant.

To accommodate time dependent road conditions in the shortest path calculation, Gonzalez in~\cite{gonzalez:tra:mining} uses a traffic mining approach
to determine the traffic patterns using the historical data for different routes at different times. Similarly, Kanoulas et al.~\cite{kanoulas:speed}
use speed patterns of previous days to compute most efficient path for a certain time.

Some techniques~\cite{mit18,mit19,mit20} rely on graph preprocessing under the assumption of static conditions (e.g., landmark) to accelerate query
response times. Reach based routing~\cite{mit18} enhances query responses by adding shortcut edges to reduced nodes' {\itshape reaches} during
preprocessing. Landmark indexing~\cite{mit19} and transit routing~\cite{mit20} boost run time query performance by using precomputed distances
between certain set of landmarks chosen according to the algorithms. An adaptation of landmark based routing  in dynamic scenarios~\cite{mit21}
yields improved query times but requires a link's cost not to drop below its initial value. A dynamic variant of highway node routing~\cite{mit22}
gives fast response times but can handle a very small number of edge weight changes.

Recently an approach for continuous route planning queries over a road network~\cite{cartel}  has been proposed which overcomes the limitations of the precomputation based algorithms. This approach proposes two classes of approximate techniques {\itshape K-paths} and {\itshape proximity measures} to speed up processing of the set of designated routes specified by continuous route planning queries in the face of incoming traffic delay
updates. Rather than recomputing on every update, this technique sends the user a new route only when delays change significantly. However to
facilitate this system a huge exterior hardware setup needed. For example to get raw GPS data 30 taxi cabs were deployed in a city. Then the raw data
gets preprocessed for identification of traversed road segments and estimation of delays.

Although in some of the above approaches the dynamic road conditions is considered, all of these existing approaches treat each user query
individually and identification of the similarities among users' queries is beyond their scope. In this paper, we propose an algorithm that considers
the group behavior of the queries and calculate the shortest path for a large number of queries. The idea is to reduce the computation cost by
grouping queries. There are several clustering algorithms~\cite{traclus}, which use properties such as density, similarity, etc. to cluster nodes,
lines and trajectories. However, the use of clustering techniques in shortest path calculation has not been addressed so far.

\section{Group Based Path Queries (GBPQ)}

We propose an efficient approach to compute \emph{group based path queries} (GBPQ) on road networks. A path query takes a source and a destination as
inputs and returns a sequence of road segments that minimizes the total travel cost (e.g., travel time) from the source to the destination. Given a
set of path queries, group based path queries $(GBPQ)$ cluster $n$ queries into groups and evaluates the group of path queries collectively.

\subsection{Intuition}

\eat{In a big city, a large number of people travel everyday from a suburb to the downtown in the morning time and return from the downtown to the
suburb in the evening. Both the travel paths and travel times for many of these commuters overlap each other, specially residents who live in the
same suburb and work in the downtown. Each user independently issues a path query to know the best travel path to reach his/her destination. Due to
the similar travel patterns of commuters, query answers of many of these users share a large portion of common routes. If each query is treated
individually, the common portions of these query answers need to be computed multiple times which increases the computational overhead significantly.
Our goal is to evaluate the queries collectively whose answers share a common route.

\begin{figure}[htbp]
    \centering
    \includegraphics[width=3.0in]{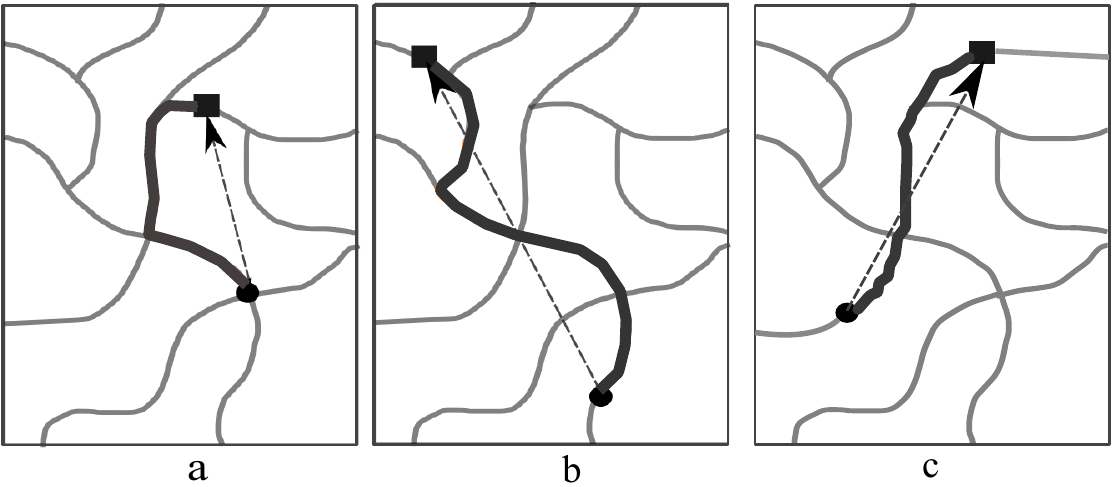}
    \caption{Pattern between source-destination trajectory and paths.}
    \label{fig:roadPatterns}
\end{figure}

}

The basic idea of our approach is developed based on a key observation of road networks, i.e., path coherence. Path coherence is a property that
shows that the shortest paths from nearby sources to nearby destinations share a large common path among them. That is, two nearby source vertices
$s_1;s_2$ will have large road segments common of their shortest paths to reach two nearby destination vertices $d_1;d_2$. Thus, if we can group
these source-destination pairs together, we can reduce the computational overhead significantly by the shared execution of these queries. To group
these source-destination pairs, we use the query line (Q-line) similarities. A Q-line is a straight line connecting a source (e.g., $s_1$) and a
destination (e.g., $d_1$). Based on the path coherence properties,  it is highly likely that queries who have similar Q-lines will have a large
portion of common travel path. Hence, we propose a group based approach that groups source and destination pairs based on their Q-lines similarities
and execute the group query on the road network.

\eat{That is, the shortest path remains close to the straight line path (i.e., Euclidian path) that connects the source and the destination of the query. We term this straight line as a query line (Q-line) in this paper. Figure~\ref{fig:roadPatterns} shows examples of some general travel patterns: (a) most of the travel paths are in one side of the Q line, (b) travel paths are distributed in the both sides of the Q-line, and (c) roads segments remain very close to the Q-line. We can easily discover these patterns in a planned cities like Burbank, California, USA (Figure~\ref{fig:burbank}~(a)) as well as in an unplanned city like Dhaka (Figure~\ref{fig:burbank}~(b)). In the figure, the bold line denotes the shortest path from a source to a destination. A source is marked with filled circle and a destination is marked with filled rectangle. The dotted connecting line between them is the Q-line of that particular query. }

It is highly likely that queries who have similar Q-lines will have a large portion of common travel path. Hence, we propose a group based approach
that groups source and destination pairs based on their Q-lines similarities and execute the group query on the road network.

\eat{
\begin{figure}[htbp]
  \begin{center}
    \begin{tabular}{cc}
         \hspace{-5mm}
      \resizebox{42mm}{!}{\includegraphics{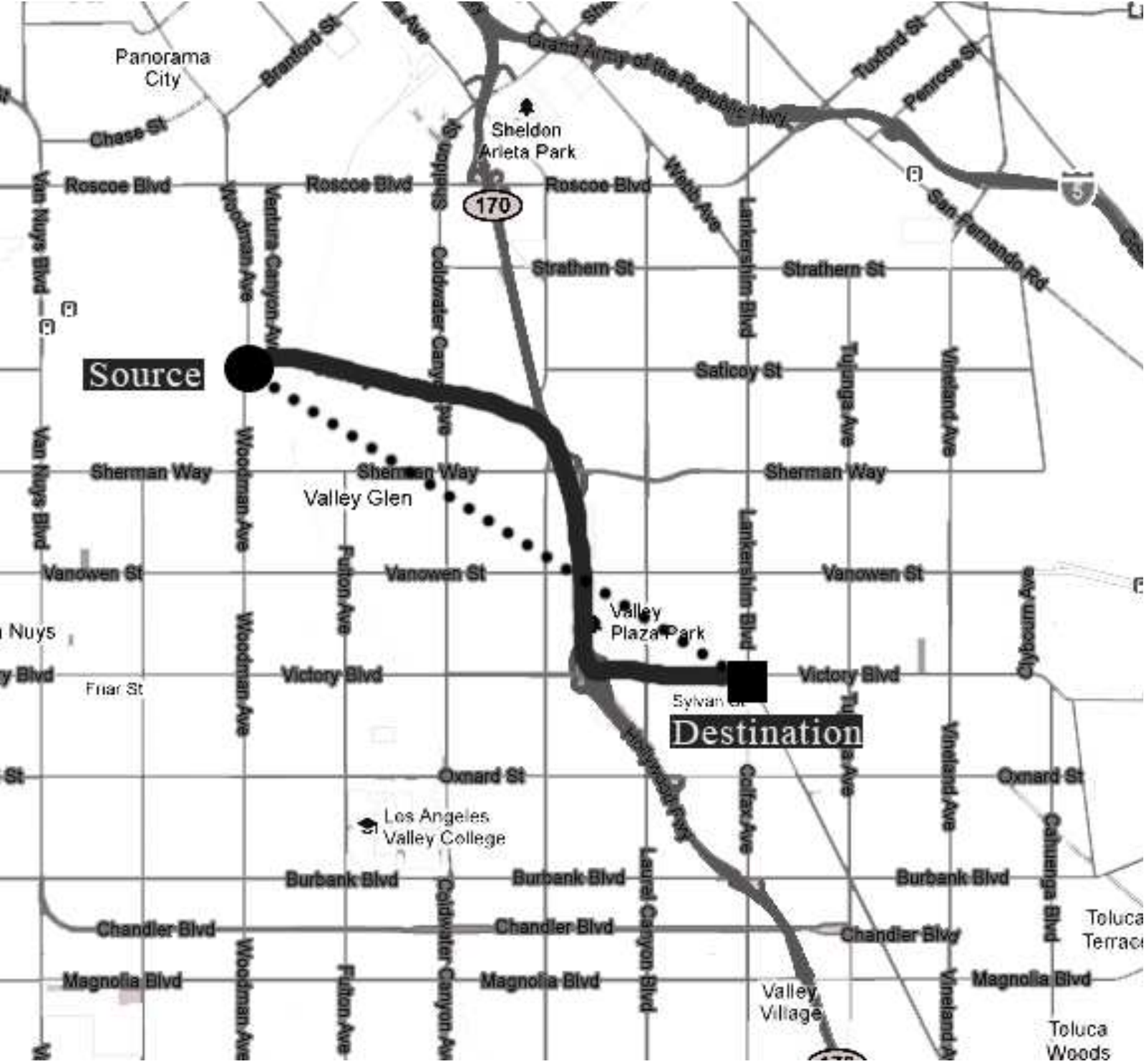}} &
         \hspace{5mm}

      \resizebox{42mm}{!}{\includegraphics{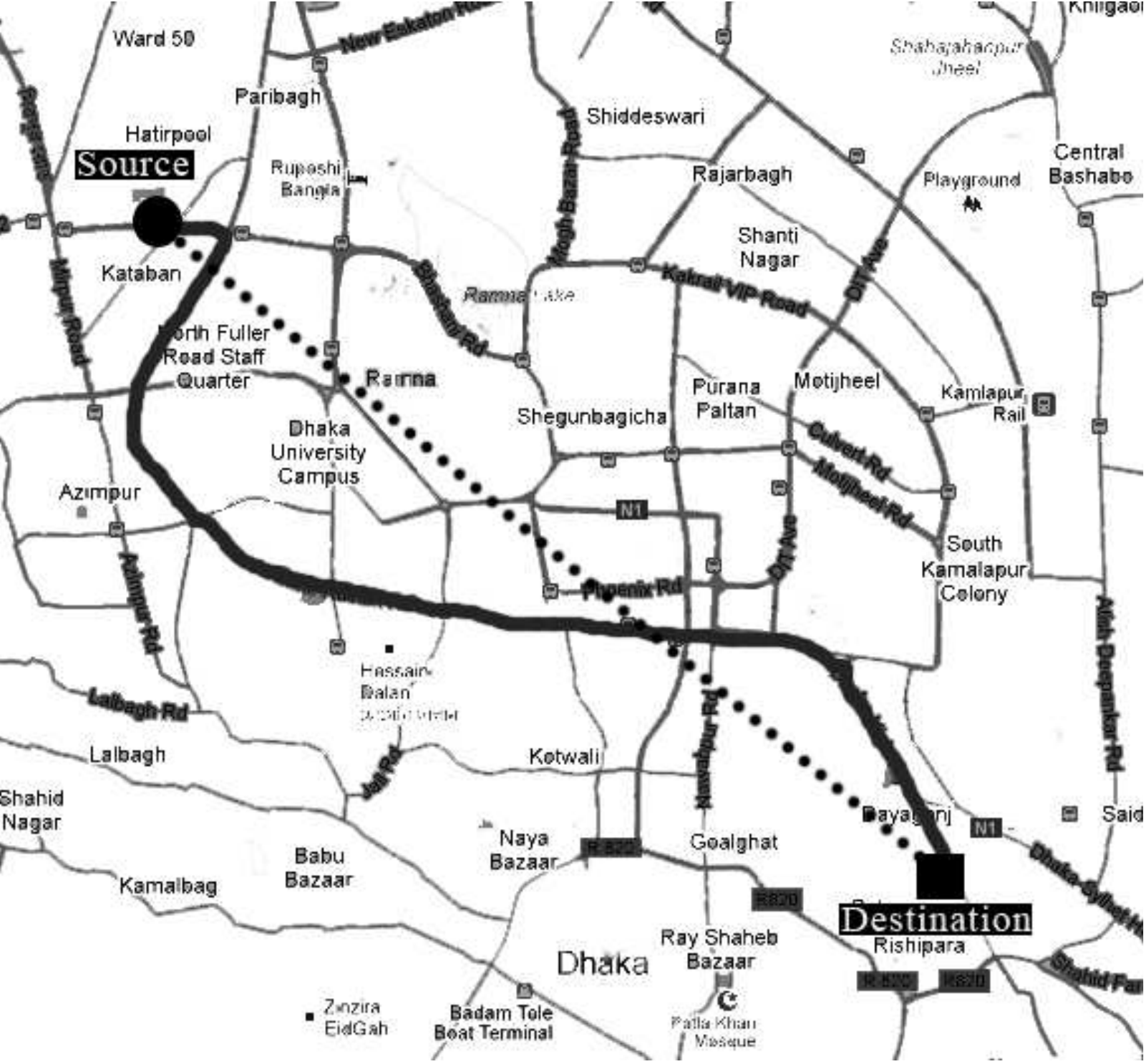}} \\
      \scriptsize{(a)\hspace{0mm}} & \scriptsize{(b)}
        \end{tabular}
   \caption{Google Maps view of (a) Burbank, California, USA and (b) Dhaka, Bangladesh.}
\label{fig:burbank}
  \end{center}
   \vspace{-5mm}
\end{figure}
\vspace{-5mm}
}

\subsection{Solution Overview}
Our approach GBPQ for processing shortest path queries groups source destination pairs into different clusters, based on the Q-lines, the connecting
lines between the source and destination pairs. Source region and destination region are obtained by combining the sources and the destinations,
respectively, of the same group. These regions are treated as the source and destination of all the points of their corresponding clusters. A region
acts like a virtual super-node having region exit paths as its edges. Next, the {\it weighted shortest path} from the source region to the
destination region is computed, where the weights refer to the travel cost (travel time, or travel length) of the path. Finally, the shortest path
for every source-destination pair that belongs a cluster is computed by concatenating three path fragments: (i) shortest path from the source point
to the starting point of the source-destination region shortest path, (ii) source region to destination region shortest path of the cluster, and
(iii) shortest path from the end point of source-destination region shortest path to the destination point.

\begin{figure}[htbp]
  \begin{center}
    \begin{tabular}{cc}
         \hspace{-2mm}
      \resizebox{58mm}{!}{\includegraphics{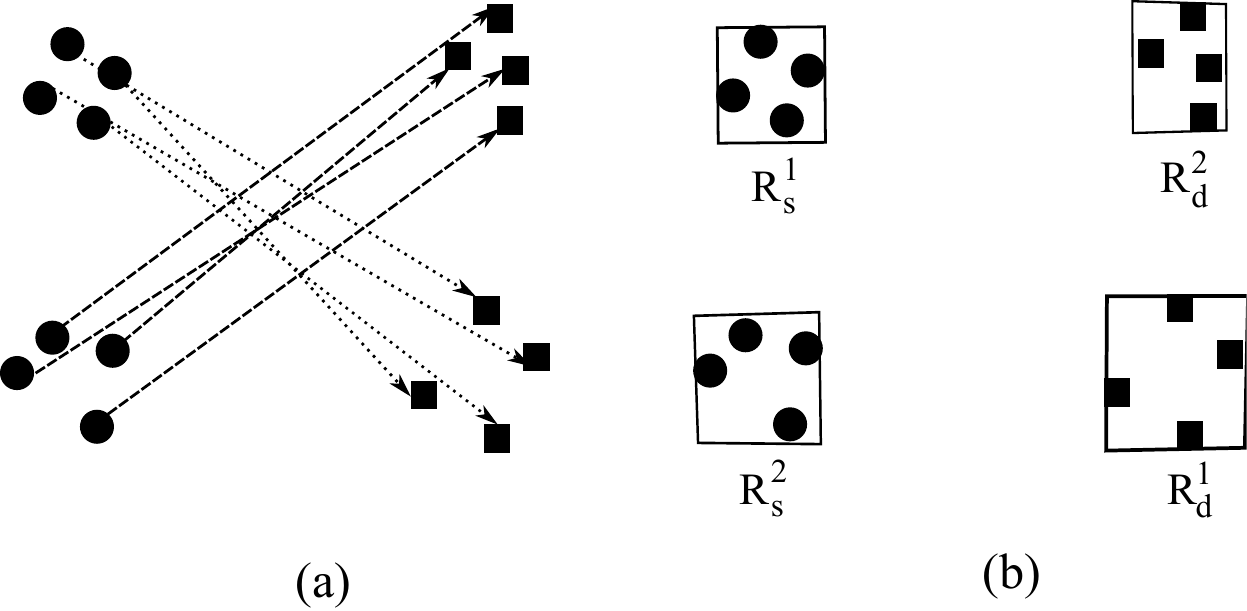}} &
         \hspace{5mm}

      \resizebox{58mm}{!}{\includegraphics{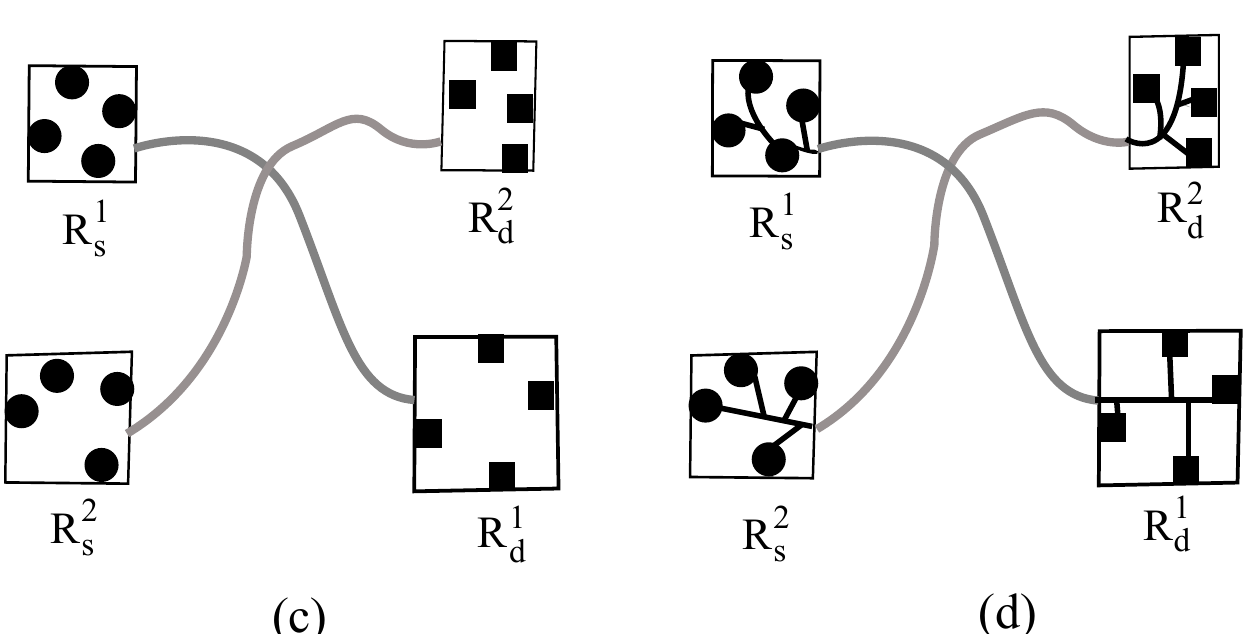}} \\
        \end{tabular}
   \caption{An example of finding shortest paths in a group based framework.}
\label{fig:steps}
  \end{center}
   \vspace{-5mm}
\end{figure}
\vspace{-5mm}

Figure~\ref{fig:steps} illustrates the procedure with eight path queries: (a) obtaining Q-lines from initial source and destination points of path
queries, (b) clustering source and destination points and creating region pairs $\{R_s^1, R_d^1\}$ and $\{R_s^2, R_d^2\}$, (c) finding shortest path
between two regions, and (d) adding up internal paths within a region to obtain the complete path.

Note that since we have considered only the best path between a source region and its corresponding destination region for a group, this path may not
be the best path for every source-destination point pairs that belong to this group. Thus individual query in GBPQ may result in slightly larger path
then the optimal shortest path. Our goal is to keep the deviation from optimal shortest path as low as possible. Our extensive evaluation shows the
deviation of the path returned by GBPQ from the optimal path is only about 0.5\% in the average case, which is within the acceptable limit of the users.

\section{Algorithm}
In this section, we present our algorithm for group-based path queries (GBPQ). The input to the algorithm is a set of $n$ path queries, $SD$ =
$\{(s_1, d_1), (s_2, d_2),$ $ \dots , (s_n, d_n)\}$, where $(s_i, d_i)$ represents a path query from a source $s_i$ to a destination $d_i$  for $1
\leq i \leq n$, and the output of the algorithm is a set of approximate shortest paths, $P$ = $\{p_1, p_2, \dots , p_n\}$, where $p_i$ represents the
approximate shortest path for the path query $(s_i, d_i)$. Though there exists algorithms~\cite{dijkstra,Nilson:Astar} that find the optimal answers
for path queries, applying those algorithms independently for each query at a time  incur high computational overhead, which causes a barrier in
answering a large number of simultaneous path queries, especially in high load conditions. Thus, we propose a shared execution strategy that group
similar queries using some key features of road networks. The algorithm first finds a common shortest path with respect to each group of path queries
and then computes the approximate shortest path for each individual path query $(s_i, d_i)$ based on the common shortest path of the group. Our
approach sacrifices the accuracy of the query answers slightly, i.e., computes a slightly larger path than the optimal one, in turn for significant
savings in computation time.

In Table~\ref{table:symbol}, we have summarized different symbols used in this section.
\vspace{-5mm}
\begin{table}[htbp]
\begin{center}
\caption{Symbols used in our approach.} \label{table:symbol}
\begin{tabular}{|p{0.8in}|p{3.5in}|}
\hline
Symbol & Description\\
\hline\hline
$SD$ & A set of $n$ path queries $\{(s_1, d_1), (s_2, d_2),$ $ \dots , (s_n, d_n)\}$\\
$L$ & A set of Q-lines $\{l_1, l_2, \dots , l_n\}$\\
$P$ & A set of approximate shortest paths $\{p_1, p_2, \dots , p_n\}$\\ \hline
$C$ & A set of clusters $\{c_1, c_2, \dots , c_m\}$\\
$R$ & A set of region pairs $\{(R_s^1, R_d^1), (R_s^2, R_d^2), \dots, (R_s^m, R_d^m) \}$\\
$SP$ & A set of weighted shortest paths $\{sp_1, sp_2, \dots, sp_m\}$\\ \hline
$\Delta$ & Half length of the side of areas of influence\\
$\psi$ & Distance threshold to group queries while forming clusters\\
$\mu$ & Minimum required number of queries to form a cluster\\
\hline
\end{tabular}
\end{center}
\end{table}
\vspace{-5mm}

\setlength{\algomargin}{2em} \dontprintsemicolon
\begin{algorithm}[htbp]
\begin{small}

\SetKwInOut{Input}{Input} \SetKwInOut{Output}{Output} \label{algo:groupBasedPathQueries} \caption{\textsc{Evaluate\_GBPQ}($SD$)}
\Input{$SD$ = $\{(s_1, d_1), (s_2, d_2), \dots , (s_n, d_n)\}$}
\Output{$P$ = $\{p_1, p_2, \dots , p_n\}$}

    \tcc{Q-line formation}
    \For {each $(s_i, d_i)$ $\in$ $SD$}
    {\label{algo:1:line:traform:begin}
        $l_i \leftarrow getStraightLine(s_i, d_i)$ \label{algo:1:line:traform:end}\;
    }
    \tcc{Query clustering and region formation}
    $C \leftarrow $ \textsc{Cluster\_Queries} ($l_1, l_2, \dots , l_n$) \label{algo:1:line:traclus:begin}\;
    Compute $\{(R_s^1, R_d^1), (R_s^2, R_d^2), \dots, (R_s^m, R_d^m) \}$\;

\label{algo:1:line:traclus:end}\;

    \tcc{Path calculation}
    \For {each $(R_s^j, R_d^j) \in R$}
    {\label{algo:1:line:fastpath:begin}
            $sp_j \leftarrow weighted Shortest Path$ $(R_s^j, R_d^j)$\;
}

    \For {each $(s_i, d_i)$ $\in$ $SD$} {
            Find $j$ such that $s_j \in R_s^j$ and $d_i \in R_d^j$\;
            $p_i \leftarrow $ \textsc{Construct\_Path} $(s_i, d_i, sp_j)$ \label{algo:1:line:fastpath:end}\;
    }

\end{small}
\end{algorithm}
\vspace{-5mm}

Algorithm~\ref{algo:groupBasedPathQueries}, $Evaluate\_GBPQ$, gives the pseudo code for processing GBPQ. The algorithm finds the set of shortest paths, $P$, in three steps: (i) Q-line formation (Lines \ref{algo:groupBasedPathQueries}.1 --~\ref{algo:groupBasedPathQueries}.2), (ii) Q-line clustering and region formation (Lines ~\ref{algo:groupBasedPathQueries}.3 --~\ref{algo:groupBasedPathQueries}.4) and (iii) path calculation (Lines~\ref{algo:groupBasedPathQueries}.5 -- ~\ref{algo:groupBasedPathQueries}.9). We discuss the details of three steps in the following sections.

\subsection{Q-line Formation}
We define Q-line, $l_i$, as the straight line connecting the source $s_i$ and the destination $d_i$ of a path query. The concept of Q-line is used to
predict the similarity of path queries, whose answers share common paths. The algorithm computes the set of Q-lines, $L$ = $\{l_1, l_2, \dots ,
l_n\}$ in Line 1.2, which is used for clustering the path queries in the second step of the algorithm.

\subsection{Query Clustering and Region Formation}

Algorithm~\ref{algo:groupBasedPathQueries} clusters the path queries based on the similarities of Q-lines and then computes the \emph{source and
destination region pair} for each cluster. A source and destination region pair consists of a source region and a destination region, where the
source region (destination region) of a cluster is a minimum bounding rectangle (MBR) containing the locations of sources (destinations) of all path
queries in the cluster. Note that we use MBRs to represent source and destination regions because the tighter the regions the less is the deviation of the computed path from the optimal one.

Algorithm~\ref{algo:groupBasedPathQueries} finds the set of clusters $C$ = $\{c_1, c_2, \dots , c_m\}$ using the function $Cluster\_Queries$ (Line
1.3), where $m \leq n$. Algorithm~\ref{algo:Q-lineGrouping} shows the steps for $Cluster\_Queries$. The detail discussion of Function
$Cluster\_Queries$ is given in Section~\ref{sec:cluster_queries}. To measure the similarity among Q-lines, we define two metrics: (i) \emph{distance
function}, and (ii) \emph{areas of influence}, which are used in Function $Cluster\_Queries$. Therefore, we first explain distance function and areas
of influence in Section~\ref{sec: distance_function} and Section~\ref{sec: area_of_influence}, respectively.

After clustering the queries, Algorithm~\ref{algo:groupBasedPathQueries} (Line 1.4) computes the set of source and destination region pairs, $R$ =
$\{(R_s^1, R_d^1), (R_s^2, R_d^2), \dots,$ $ (R_s^m, R_d^m) \}$, where $R_s^j$ and $R_d^j$ represent source region and destination region,
respectively, of cluster $c_j$ for $1 \leq j \leq m$.

\subsubsection{Distance function:}
\label{sec: distance_function} 
\begin{figure}[htbp]
  \begin{center}
    \begin{tabular}{cc}
         \hspace{-5mm}
      \resizebox{46mm}{!}{\includegraphics{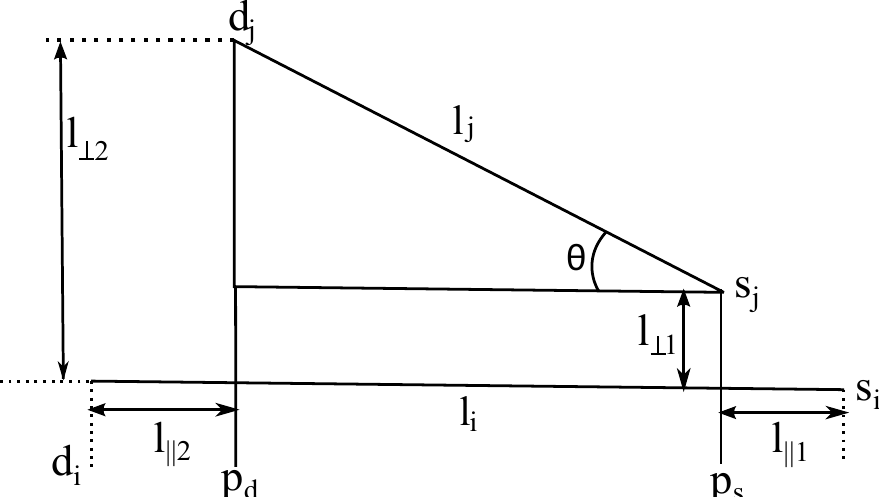}} &
         \hspace{5mm}
      \resizebox{44mm}{!}{\includegraphics{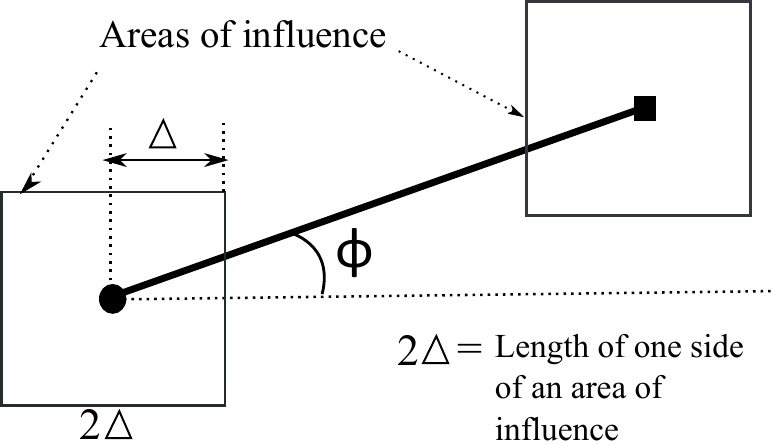}} \\
      \scriptsize{(a)\hspace{0mm}} & \scriptsize{(b)}
        \end{tabular}
   \caption{(a) Components of the distance functions, (b) Influence area of a Q-line}
\label{fig:distances}
  \end{center}
   \vspace{-5mm}
\end{figure}

A distance function is used to measure the similarities among Q-lines of user queries. We use three distance measures
and combine them to measure the distance between any two Q-lines. Three distance measures are: (i) parallel distance $d_\parallel$, (ii)
perpendicular distance $d_\perp$ and (iii) angular distance $d_\theta$. We calculate the distance between two Q-lines by using the following formula:
\begin{equation}
distance = w_\perp d_\perp + w_\parallel d_\parallel + w_\theta d_\theta
\end{equation}
The weight values $w_\perp$, $w_\parallel$ and $w_\theta$ are used to control the effective contribution of three components on the overall distance.
For example, a larger $w_\parallel$ value reduces the length difference between a Q-line and its projection on the other Q-line. Similarly a larger
value of $w_\perp$  keeps the endpoints of two queries closer to each other. In our experiments, we keep all these weights to unity, so that all the
three components of the distance function have equal effect on overall distance. We group user queries based on the distance function. Two path
queries can be grouped together if their distance is less than a threshold value $\psi$. The formal definitions and impact of parallel, perpendicular
and angular distances~\cite{traclus} are discussed below. Symbols used in the definitions are shown in Figure~\ref{fig:distances}a.


\textsc{Parallel Distance:} Let $s_j$ and $d_j$ be the two endpoints of the Q-line $l_j$. If the projection of $s_j$$d_j$ over $l_i$ is $p_s$$p_d$,
then the parallel distance is defined as maximum of the Euclidean distance of $s_i$ to $p_s$ and $d_i$ to $p_d$ as shown in
Equation~\ref{egn:paralleldistance}.
\begin{equation}
\label{egn:paralleldistance} d_\parallel =MAX(s_ip_s, p_dd_i)
\end{equation}

\textsc{Perpendicular Distance:} Let $l_{\perp1}$ and $l_{\perp2}$ be the distance components of two Q-lines $l_i$ and $l_j$ as shown in
Figure~\ref{fig:distances}. Then the perpendicular distance of these two Q-lines is defined with second order Lehmer mean~\cite{lehmermean} of
$l_{\perp1}$ and $l_{\perp2}$ as shown in following equation.
\begin{equation}
\label{egn:perpdistance} d_\perp = \frac{ l^2_{\perp1} + l^2_{\perp2} }{ l_{\perp1} + l_{\perp2}}
\end{equation}

\textsc{Angular Distance:} Let $\theta$ be the smaller intersecting angle between Q-lines $l_i$ and $l_j$. Then their angular distance is defined as
product of $sin$ component of the larger Q-line. Mathematically defined as following equation.
\begin{equation}
\label{egn:angulardistance} d_\theta = MAX\{l_i,l_j\}\times sin\theta
\end{equation}

In summary, a smaller value of $d_{\parallel}$ ensures that the difference between the length of one Q-line and the length of the projection of another Q-line over the first one to be smaller. On the other hand, with the increase of the value of $d_{\perp}$, the endpoints of two Q-lines move farther away from each other. So when both of these distances, $d_{\parallel}$ and $d_{\perp}$, proceed towards zero, query locations get close to each other and query distances, the straight line distance between a source and a destination, converge towards equality. Finally, the angular distance $d_{\theta}$ checks the parallelism between two Q-lines. When $\theta = 0$ two Q-lines are parallel to each other. Thus, for exactly similar two Q-lines, the overall distance value is zero.


For clustering queries, similarities among Q-lines are measured using the distance function. However, computing the similarities for every pair of
Q-lines will be prohibitively expensive, specially for a large number of Q-lines. Moreover, it is unnecessary to compute the similarities between two
Q-lines if they are far apart from each other. Thus, to decide on which Q-lines should be compared with each other, we introduce the concept of
{\itshape areas of influence}. The areas of influence helps to prune a large number of Q-lines while forming clusters' of queries and thus reduces
the computational overhead significantly.

\subsubsection{Areas of influence:}
\label{sec: area_of_influence} We define the areas of influence of a Q-line as a pair of regions in which, if another Q-line is present, there is a high probability that the distance between those two Q-lines is smaller than the threshold value $\psi$. These regions are represented as  two squares centering the two endpoints of the Q-line. Now, if another Q-line has its source and destination inside these two squares, respectively, then we say the later Q-line is in the  same region/area with the first one and compute the distance between these two Q-lines to check whether they belong to the same cluster.

We define the side length of both squares as $2\Delta$. Thus the value of $\Delta$ defines the size of the areas of influence. A larger value of $\Delta$ results in many unwanted Q-lines to be included for distance calculations and a smaller value results in a large number of small clusters. Since the appropriate value of $\Delta$ depends on the query set, we choose a $\Delta$ through a empirical study in the experiment. We find a suitable Figure~\ref{fig:distances}b shows the properties of areas of influence. There are various other options to choose the type of the region such as rectangle, circle or ellipse. Our approach is independent to the shape of the influence area, however, a proper distance function may need to be developed for a chosen shape.

The concepts of distance function and areas of influence form the bases of our algorithm {\itshape Cluster\_Queries}.

\subsubsection{Cluster queries:}
\label{sec:cluster_queries} The input to Algorithm~\ref{algo:Q-lineGrouping} is the set of Q-lines, $L$~=~$\{l_1, l_2,$ $\dots, l_n\}$. We take each
Q-line $l$ from $L$ and find the set of Q-lines  $I$  that are inside $l$'s areas of influence (Line~\ref{algo:Q-lineGrouping}.4). The initial
{\it representative Q-line} , $r$, of the set $I$ is computed in Line~\ref{algo:Q-lineGrouping}.5. The {\it representative Q-line} of a given set of
Q-lines is the average direction vector of those Q-lines. Average direction vector $\bar{V}$ of the set $V$ = $\{v_1, v_2, \dots , v_n\}$ is
calculated with the following equation: $ \bar{V}= \frac{v_1+ v_2+ \dots + v_n}{|V|} $, where $|V|$ is the cardinality of $V$.

Now, for every element of $I$, we first calculate its distance with the current representative Q-line. If the value is less than or equal to $\psi$,
we take this Q-line to be added to the new cluster. The representative Q-line $r$ is then updated to show the effect of newly added Q-line
(Lines~\ref{algo:Q-lineGrouping}.6 -~\ref{algo:Q-lineGrouping}.10). We use a moving average process to update $r$. For example, if $r$ is
representing $n$ queries of a cluster, then when a new Q-line $i$ is added into the cluster the new value of $r$ will be updated as $\frac{r * n + i
}{n+1}$. We take moving average because in some cases it may happen that the set $I$ contains Q-lines that should have been clustered in different
two groups. However, as the initial representative Q-line is the average of all Q-lines in the set, Q-lines of both the probable clusters may have
lesser distance than $\psi$. When $r$ is updated with the  moving average of clustered Q-lines, chance of $r$ shifting towards one cluster increases,
depending on the sequence of Q-lines that are updating $r$. In the worst case, if Q-lines form both probable clusters  update $r$ alternatively,
the improvement might be very low. The Q-lines included in the new cluster are removed from $L$.

\setlength{\algomargin}{2em} \dontprintsemicolon
\begin{algorithm}[H]
\begin{small}
\SetKwInOut{Input}{Input} \SetKwInOut{Output}{Output} \label{algo:Q-lineGrouping} \caption{\textsc{Cluster\_Queries}($L$)}
\Input{A set of Q-line $L$ = $\{l_1, l_2, \dots , l_n\}$}
\Output{A set of clusters of queries $C$ = $\{c_1, c_2, \dots , c_m)\}$}

    $L^\prime \leftarrow Null$ \tcp*[r]{backup list}\;
    \For {each $l$ $\in$ $L$}
    {
         $Q \leftarrow Null$ \tcp*[r]{list for current cluster}\;

         $I \leftarrow subset Within Area Of Influence(L,\; l)$ \label{algo:cluster:line:infset}\;

         $r \leftarrow representative Q$-$line(I)$ \label{algo:cluster:line:repline}\;
         \For{each $i$ $\in$ $I$}
         { \label{algo:cluster:line:cluster}
              \uIf {$ distance(r, i) \leq \psi $}
              {
                  $Q \leftarrow i$\;
          Update(r)\;
                  Remove $i$ from $L$ \label{algo:cluster:line:cluster:end}\;
               }
         }
         \uIf {$size(Q) \leq \mu$}
         {\label{algo:cluster:line:create}
            $Create\_Cluster(Q, r)$ \;
         }\uElse
         {
            $L^\prime \leftarrow Q$ \label{algo:cluster:line:create:end}\;
          }
    }

         \For {each $q$ $\in$ $L^\prime \cup L$}
         { \label{algo:cluster:line:remain}
            \For {each $c $$\in$$ C$}
            {
                 \uIf {$ distance(representative Q$-$line(c), q) \leq \psi \; \textbf{and}$ q is not classified}
                 {
                       $c \leftarrow q$\;
                       Update(r)\;
                       Mark q as classified\;
                 }
            }
                 \uIf{q is not classified} {
                      $Create\_Cluster(\{q\}, q)$\;
                 }
        Remove $q$ from $L^\prime$ or $L$ \label{algo:cluster:line:remain:end}\;
       }

\end{small}
\end{algorithm}

However, if the distance value is greater than $\psi$ we simply discard that Q-line. Also, a new cluster is only created if there are at least $\mu$
number of Q-lines present. If there are not sufficient number of Q-lines then we put them in a back up list $L^\prime$ and classify them later
(Lines~\ref{algo:Q-lineGrouping}.12 -~\ref{algo:Q-lineGrouping}.15).


When the initial clustering process is completed, we have some Q-lines which are not included in any cluster. Some Q-lines are not clustered because the representative
Q-line was initially at a greater distance than $\psi$ and later it did not fall into others' influence areas, i.e., the remaining elements of the
set $L$. The other set of Q-lines $L^\prime$ may be left unclustered because there are not sufficient amount of queries to form a new cluster
(Line~\ref{algo:Q-lineGrouping}.15). For all these queries, we initially check them with already created clusters. If we find any cluster with
lesser distance than $\psi$, we add the query into that cluster and update its representative Q-line. Otherwise a new cluster is created and the
Q-line is removed from $L$ or $L^\prime$ (Lines~\ref{algo:Q-lineGrouping}.17 -~\ref{algo:Q-lineGrouping}.26).

\subsection{Path Calculation}
\begin{figure}[htbp]
  \begin{center}
    \begin{tabular}{cc}
         \hspace{-5mm}
      \resizebox{42mm}{!}{\includegraphics{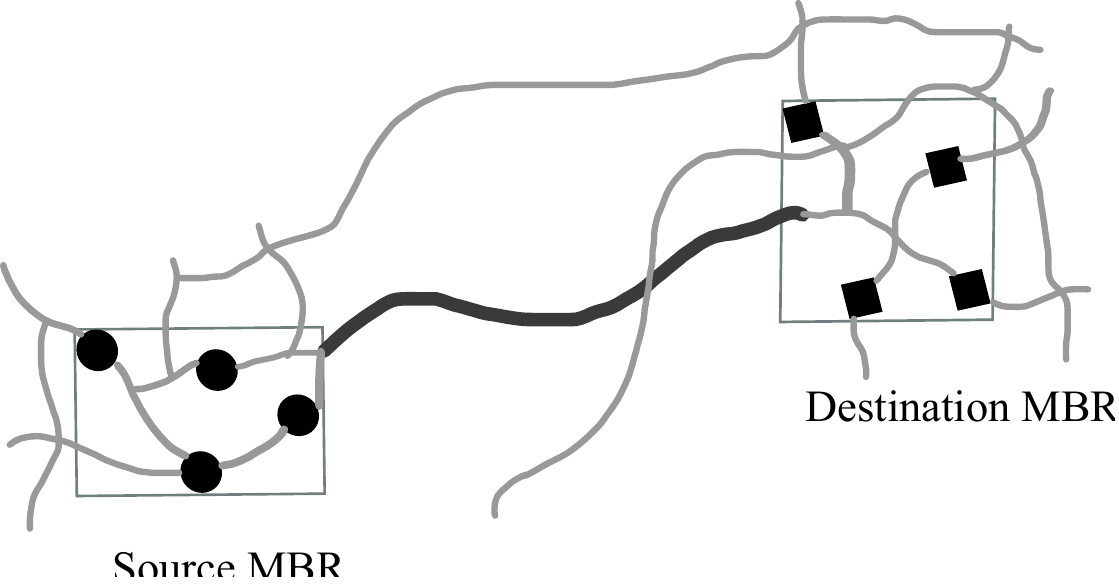}} &
         \hspace{5mm}

      \resizebox{42mm}{!}{\includegraphics{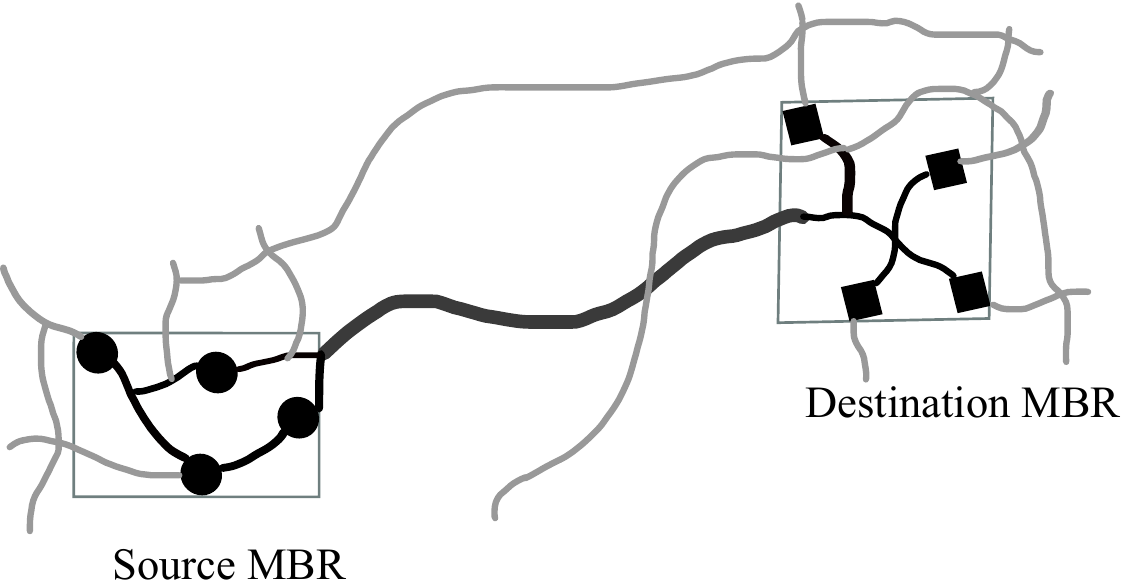}} \\
      \scriptsize{(a)\hspace{0mm}} & \scriptsize{(b)}
        \end{tabular}
   \caption{(a) Finding the weighted shortest path, (b) Constructing shortest path by adding road segments within regions.}
\label{fig:pathCalc}
  \end{center}
   \vspace{-5mm}
\end{figure}

The final step of Algorithm~\ref{algo:groupBasedPathQueries} is to compute the shortest path for every path query, which is done in two phases. The
algorithm first computes the weighted shortest path for each cluster for its  source-destination region pairs as shown in Figure~\ref{fig:pathCalc}a
(Line 1.6). Then the algorithm finds the approximate shortest path for each individual path query in the cluster using Function $Construct\_Path$ as
shown in Figure~\ref{fig:pathCalc}b (Lines 1.8-1.9).

The first phase, i.e., region to region shortest path is computed as follows. Essentially, as discussed earlier, a region pair consists of two MBRs, a
source region MBR and a destination region MBR. For a source region, exit points of the region are identified by considering the outgoing edges of
the region. On the other hand, for a destination region, entry points are identified by considering the incoming edges to the region. These regions
act like virtual super-nodes, where exit ( or entry) paths from the regions are the edges of those nodes. Then we apply an heuristic based approach,
A* search algorithm, to find the shortest path between a source region to a destination region.


\vspace{-5mm}
\vspace{-3mm}

The second phase of connecting source and destination points to the corresponding region's shortest path is computed as follows. We use simple
\textsc{Construct\_Path} procedure to find the road segments within a region. For each query $(s_i, d_i)$, two path fragments are computed. One at its source
region from source point $s_i$ to the start point of the path $sp_i$. Another is at destination region from the end point
of the path $sp_i$ in $R_d^i$ to  destination point $d_i$. Here $sp_i$ is the weighted shortest path between source and destination regions.
Algorithm~\ref{algo:approxFastestPath} summarizes the process.

%

\setlength{\algomargin}{2em} \dontprintsemicolon
\begin{algorithm}[htbp]
\begin{small}
\SetKwInOut{Input}{Input} \SetKwInOut{Output}{Output} \label{algo:approxFastestPath} \caption{\textsc{Construct\_Path} $(s_i, d_i, sp_j)$}
\Input{A source $s_i$, a destination $d_i$, weighted shortest path $sp_j$}
\Output{A path $p_i$}

$f_1$ $\leftarrow$ shortest path between $s_i$ and start point of $sp_j$\;
$f_2$ $\leftarrow$ shortest path between $d_i$ and end point of $sp_j$\;
$p_i \leftarrow f_1+ sp_j + f_2$\; \Return $p_i$\;
\end{small}
\end{algorithm}

\subsection{Discussion}
So far we have assumed that there are $n$ submitted queries in the system and then we apply our clustering technique to group these $n$ queries into different subgroups . The value of $n$ can be determined by the user given threshold, i.e., how long a user can wait for the query answer. Further optimization such as re-using the cluster for future incoming queries in the system is the scope of future study of the paper.

\eat{To handle incoming queries we propose two approaches. First, for those queries that arrives during the computation we wait a threshold time and starts another concurrent operation to find shortest path them. However if former computation is finished before the threshold time is elapsed we can handle these queries same as those that arrives after the computation. In this second approach, we first search for a exact match of the incoming query in our system. If the system does contain a match then stored result is returned. Otherwise we compute distance between all existing cluster i.e. their representative Q-lines and the incoming query. The incoming query is then added to a cluster that has minimum distance value among all the distance values having lesser distance than $\psi$. However, if all the queries have greater distance than $\psi$ then the incoming query is used to create an new cluster of its own.

The next issue to consider is how long do we keep the queries before discarding them from our system. Our proposal is that rather than removing the queries immediately after it is served, we keep them for a certain period of time to get better result for the incoming queries. System may recreate the clusters periodically after a given interval.}

 Lastly, we compute the shortest (or fastest) path between a source region and its destination region. Every query in that source region uses the same path to reach its destination. This may causes congestion of the traffic on that route. Also that particular path might not be optimal for all the queries of that region. To overcome this problem, an alternative approach can be to calculate $k$ shortest paths instead of only one path for each source-destination pair. Then each individual query in those region can use the best one for which the travel time is minimum. This will result in accurate paths but overall processing time will be slightly higher than GBPQ. The detailed study of using the $k$ shortest path in evaluating group based path queries is the scope of our future research.

\section{Experimental Study}
\label{chp:exper}

In this section we evaluate the performance of our proposed algorithm by varying a wide range of parameters. We compare our group based path queries
(GBPQ) approach with the naive approach that executes each query individually using A* algorithm~\cite{Nilson:Astar}. We simulate our experiment on a
system with Intel core i5 2.67 GHz processor and 4 GB of memory running Windows 7 ultimate. The language C++ is used to implement our algorithms.

A road network dataset of North America with 175,813 nodes, 179,179 edges, and a diameter of 18,579 units is used. At the beginning of experiments,
the entire map data is loaded into the memory. For a single path query we need to select a source-destination pair on the map. However to simulate a
group behavior we first partition the entire data space into a number of square windows. Then, we choose two random windows, one as a source region
and the other as a destination region for a group of path queries, who have their source locations and destination locations inside the source and
destination regions, respectively. Within a region, query points are generated using two distribution, Gaussian distribution and Zipf distribution. The effect of
distribution increases with the size of windows. For example, when the window size is 10000x10000, there is no partition as there is only one window
covering the whole data space. In such a case we actually have all the queries distributed in the whole map by using either Gaussian distribution or
Zipf distribution. Since we consider Euclidian space while generating queries, i.e, source-destination pairs, we map each point location
(source/destination) to the nearest node of the road network.

\vspace{-5mm}
\begin{table}[htbp]
\begin{center}
\caption{List to parameters}
\label{table:criteria}
\begin{tabular}{|p{2.25in}|p{0.75in}|p{0.7in}|p{0.7in}|}
\hline
{\bf Description and Symbol} & {\bf  Range}& {\bf Step Size}& {\bf  Default Value}\\
\hline\hline
Number of queries $n$ (x 1000) & 10 -- 100 & 10 & 50\\ \hline
Window length $\omega$ & 100 -- 10000& & 100\\ \hline 
Minimum query distance coefficient $d_c$ & 0.1 -- 1.0 & 0.1& 0.4\\ \hline

\end{tabular}
\end{center}
\end{table}
\vspace{-10mm}

\subsection{Parameter Tuning}

We use three pruning parameters in our experiment: i) half length of an areas of influence $\Delta$, ii) minimum distance threshold $\psi$, iii)
minimum number of queries $\mu$. Half length of an areas of influence defines two surrounding regions around source and destination points. Queries
in the same areas of influence have higher probabilities of belonging to the same cluster. Moreover, as per our distance measure, the maximum
allowable distance between two Q-lines increases with the increase of $\Delta$. The performance of GBPQ also depends on the distance threshold value
$\psi$. The parameters $\psi$ and $\Delta$ are also correlated as a higher value of $\psi$ allows us to take a larger $\Delta$. $\Delta$ is used to select the initial subset selection and then $\psi$ determines which of these queries is used to create a cluster. The other parameter $\mu$ determines the minimum cluster size, and thus has an impact on the number of clusters.
However, choosing effective values of these parameters is a challenge. Thus, we resort to detailed experimental study to choose appropriate values of
these parameters.

For a sample set of 50000 queries (1000 clusters x 50 queries each) we find different performance measures such as processing time, number of
clusters and the number of initially unclustered queries. Findings of these experiments are discussed bellow.

\subsubsection{Effect of half length of an influence area $\Delta$ and distance threshold $\psi$:}

We vary $\Delta$ from $70$ to $130$ with a step size of $10$ units. Thus, for this range of $\Delta$ values, the influence region will have a length
range from 140 to 260. In this set of experiments, our queries are generated in clusters having the window size of 100 units. Considering the size of the clusters we created we can estimate expected range of $\Delta$ and $\psi$. We have created the clusters in $100$ sq. units windows, so to inscribe that square in another square we need, in the worst case, a square with $100\sqrt{2}$ sq. units. Our expected $2\Delta$ should be close to this value. However, as it is not possible to determine the center of windows or the locations of the partitions we may need slightly larger $2\Delta$ than this. For $\psi$, consider a query having one same endpoints and for other ends one at the center and another at the corner of the respective window. In this case, queries may have distance ranges from 100 $(50+50)$ to 170 $(50+50+50\sqrt{2})$ which gives the initial intuition about the range. Number of cluster created for different values of this parameters are observed.

\begin{figure}[htbp]
\begin{center}
\begin{tabular}{cccc}
\hspace{-5mm}
\resizebox{31mm}{!}{\includegraphics{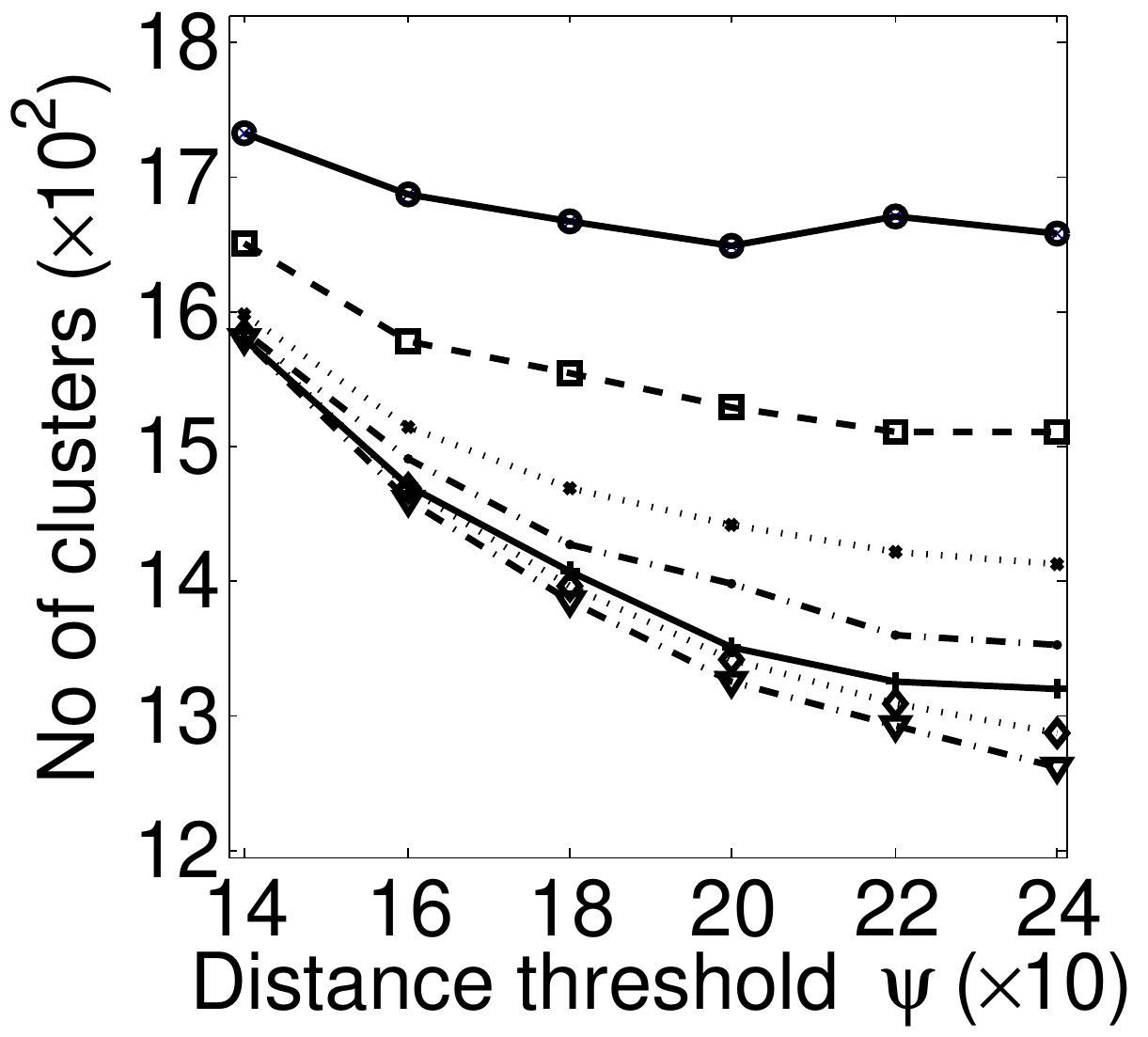}} &
\hspace{-2mm}
\resizebox{30mm}{!}{\includegraphics{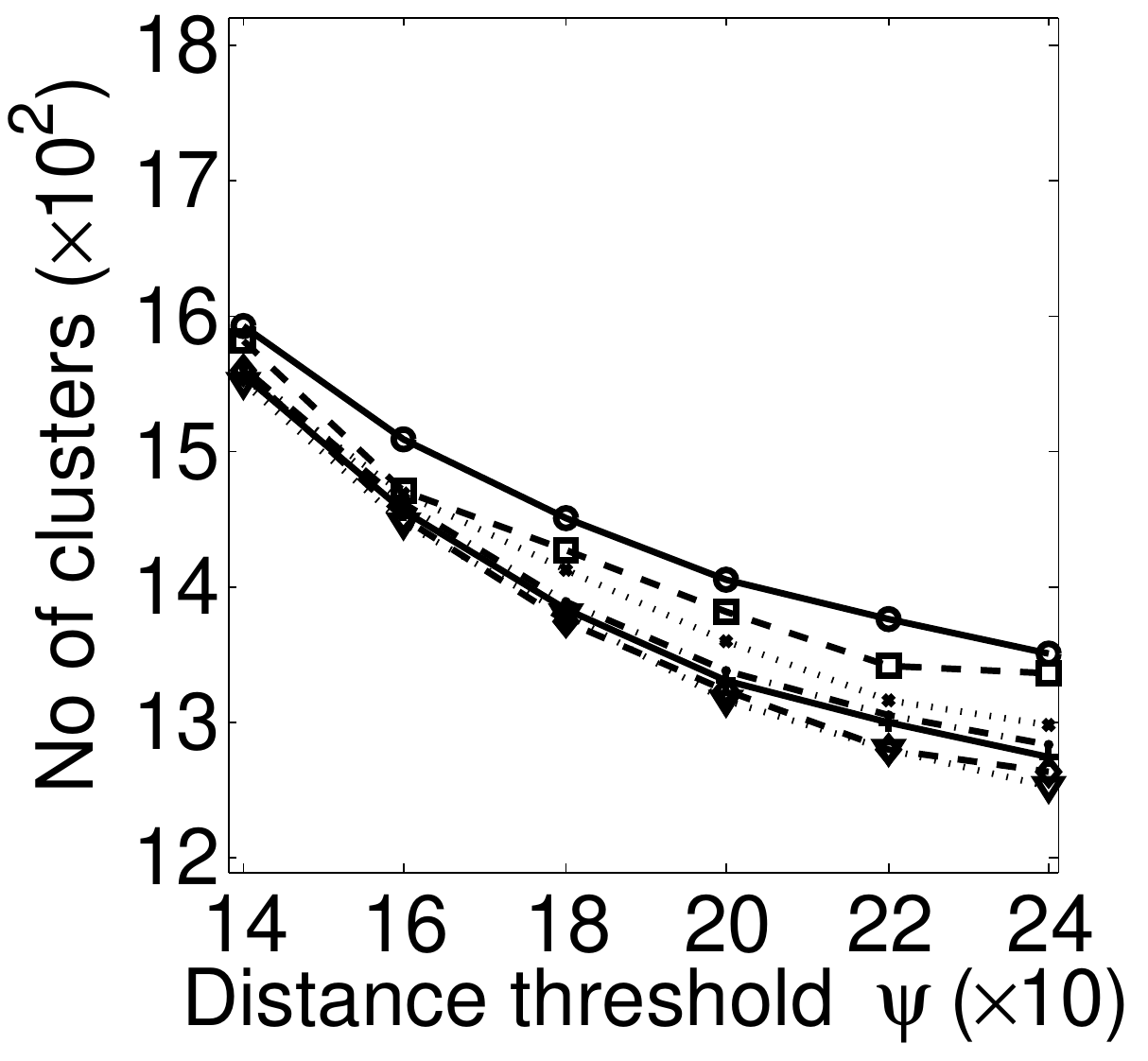}} &

\hspace{-2mm}
\resizebox{30mm}{!}{\includegraphics{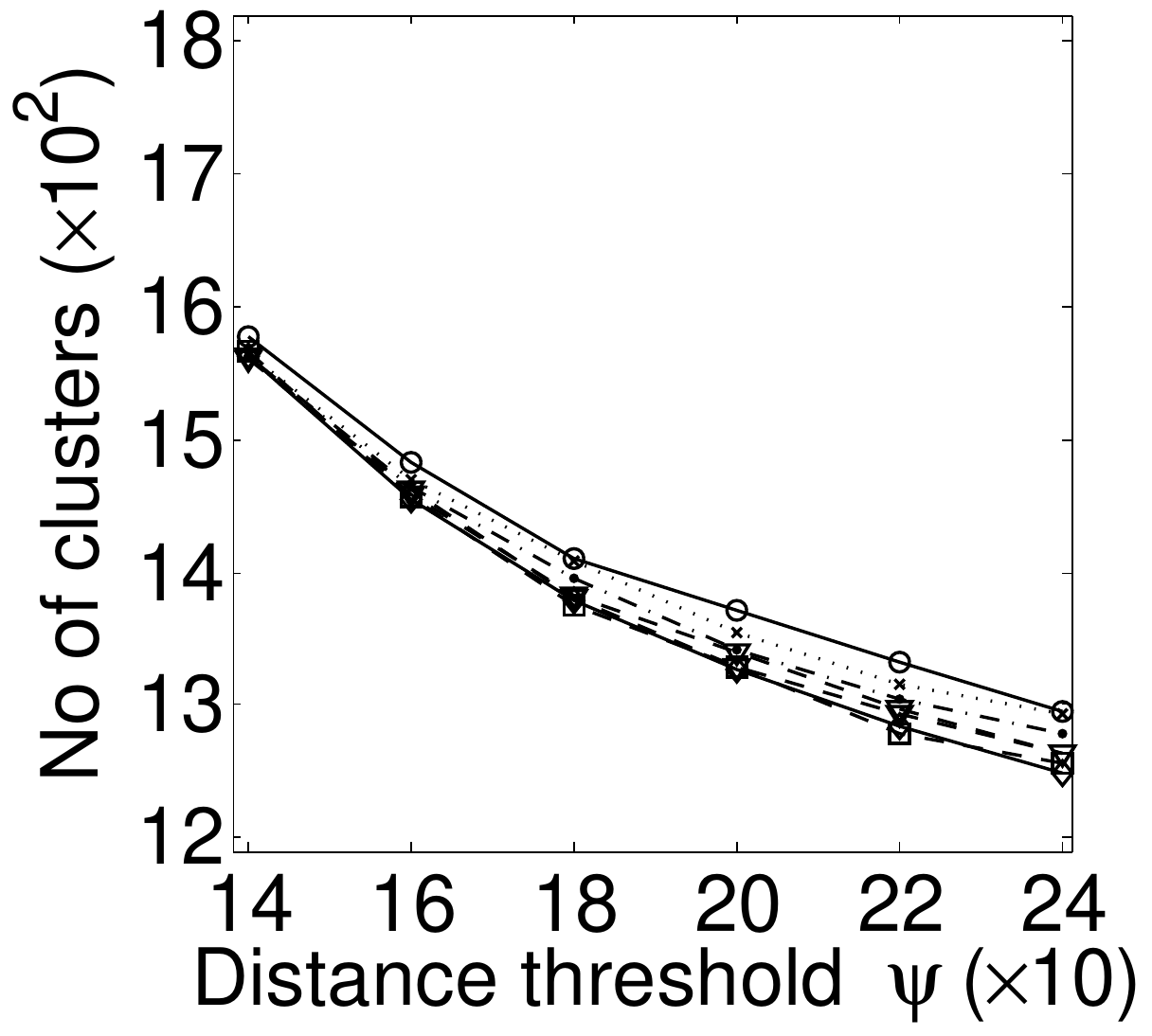}} &
\hspace{-2mm}
\resizebox{34mm}{!}{\includegraphics{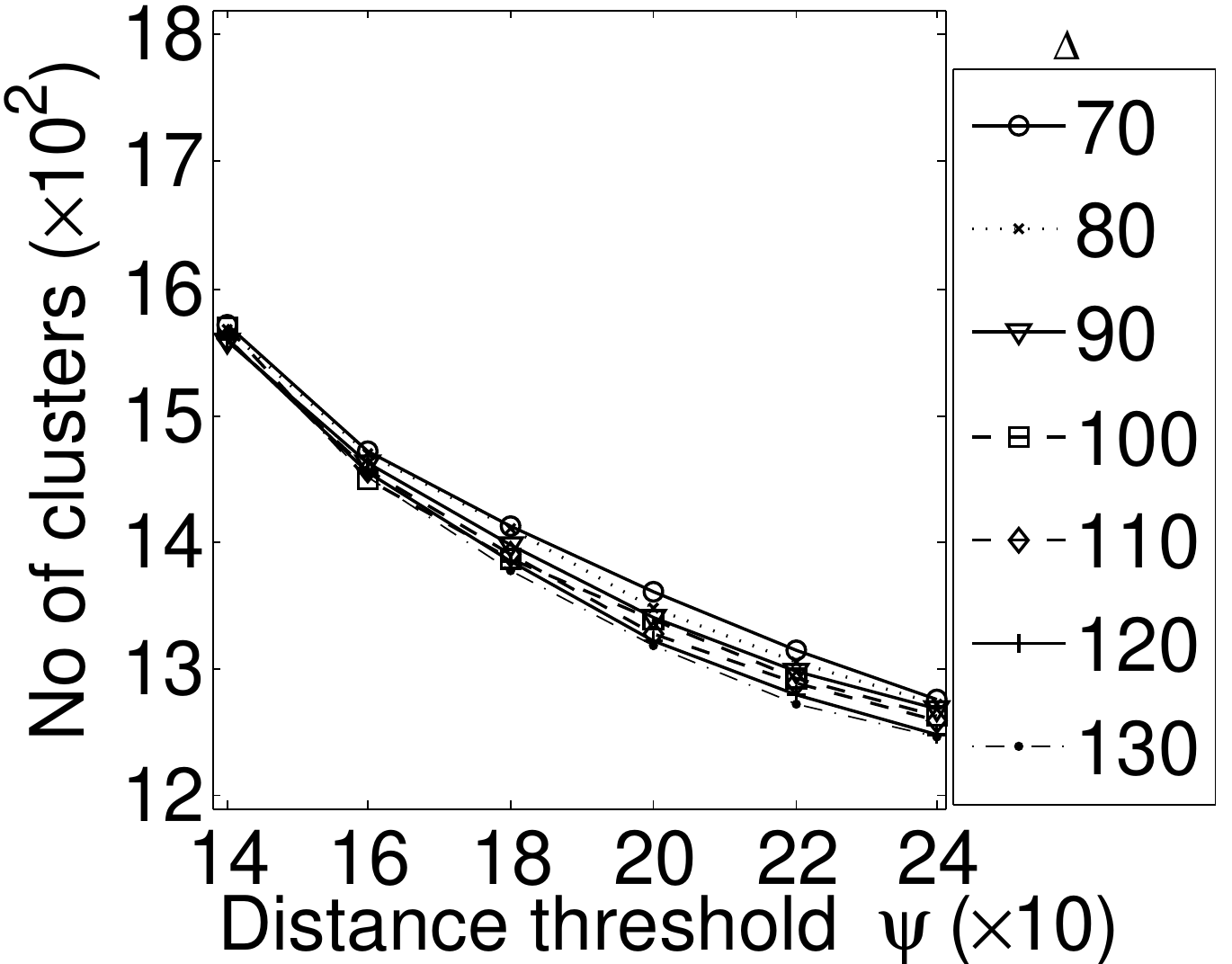}} \\
\scriptsize{(a)\hspace{0mm}} & \scriptsize{(b)} & \scriptsize{(c)\hspace{0mm}} & \scriptsize{(d)}
\end{tabular}
\caption{Effect of $\Delta$ and $\psi$ when at least (a) 01 (b) 10 (c) 20 (d) 30 queries is required to form a new cluster.} \label{fig:psi:vs:c}
\end{center}
\vspace{-5mm}
\end{figure}

Figures~\ref{fig:psi:vs:c}a, \ref{fig:psi:vs:c}b, \ref{fig:psi:vs:c}c, \ref{fig:psi:vs:c}d show the number of clusters created by varying the values
of $\Delta$ and $\psi$, when $\mu$ is fixed at 1, 10, 20, 30, respectively. The time required for the clustering is on an average 6 seconds.

Figure~\ref{fig:psi:vs:c} shows that for higher values of $\Delta$ the number of clustered decreases. The reason is that when we select a greater
influence area, more queries fall in the same area, which makes the average number of queries per region greater than that of a smaller influence
areas. For a fixed $\Delta$, when we increase the value of $\psi$, the number of clusters decreases.

\vspace{-5mm}
\vspace{-5mm}

We also compute the number of initially unclustered queries. Here, Q-lines of queries that are not clustered in the first phase of our
Algorithm~\ref{algo:Q-lineGrouping} (Lines~\ref{algo:Q-lineGrouping}.6 - \ref{algo:Q-lineGrouping}.15) is called initially unclustered
queries. The number of initially unclustered queries increases when the value of {\it $\mu$} increases, and it decreases slightly with the increasing
values of $\Delta$. For example, when $\Delta= 80$ and $\psi = 160$ the number of unclustered queries are 0, 1416, 3428, 5667, 9267, for $\mu$ values
of 0, 10, 20, 30, 40, respectively. But for $\Delta= 130$ and $\psi = 160$ the number of unclustered queries are 107, 1150, 2922, 4243, 7758,
respectively. When the value of $\Delta$ is small there are less unwanted queries (that should belong to another cluster) in the cluster but the
increase of the value of $\mu$ leaves some queries initially unclustered.

With the increase of $\Delta$, the probability of fulfilling the constraint on the minimum number of queries $\mu$ increases. However, with the
increase of $\Delta$ the number of unwanted queries in a cluster also increases. Thus, the value of $\Delta$ should not be chosen too high or too
low. Experimental results (Figure~\ref{fig:psi:vs:c}) validate this claim. We can see from the figure that $80$ should be chosen as default of
$\Delta$ as it gives consistent performance while varying other parameters.

From Figure~\ref{fig:psi:vs:c}, we can see that the graphs are more stable for $\mu$ values of 20 and 30. In these cases, for $\psi$ values of 140
and 160, the variation due to different values of $\Delta$ is insignificant. Since, the number of clusters for $\psi=160$ is smaller than that of
$\psi=160$, we can choose $160$ as our default value for $\psi$. Also, since for different values of  $\Delta$  and  $\psi$,  the range of $\mu$ between 20 and 30 gives the best performance, we can choose any value in this range as the default value of  $\mu$.

\eat{
In real world, we do not have window size. In reality window size is the size of expected clusters which can be estimated based on typical area/ block/ street etc. size.}

\subsubsection{Effect of minimum number of queries $\mu$:}
\begin{figure}[h]
\centering
\includegraphics[width=40mm]{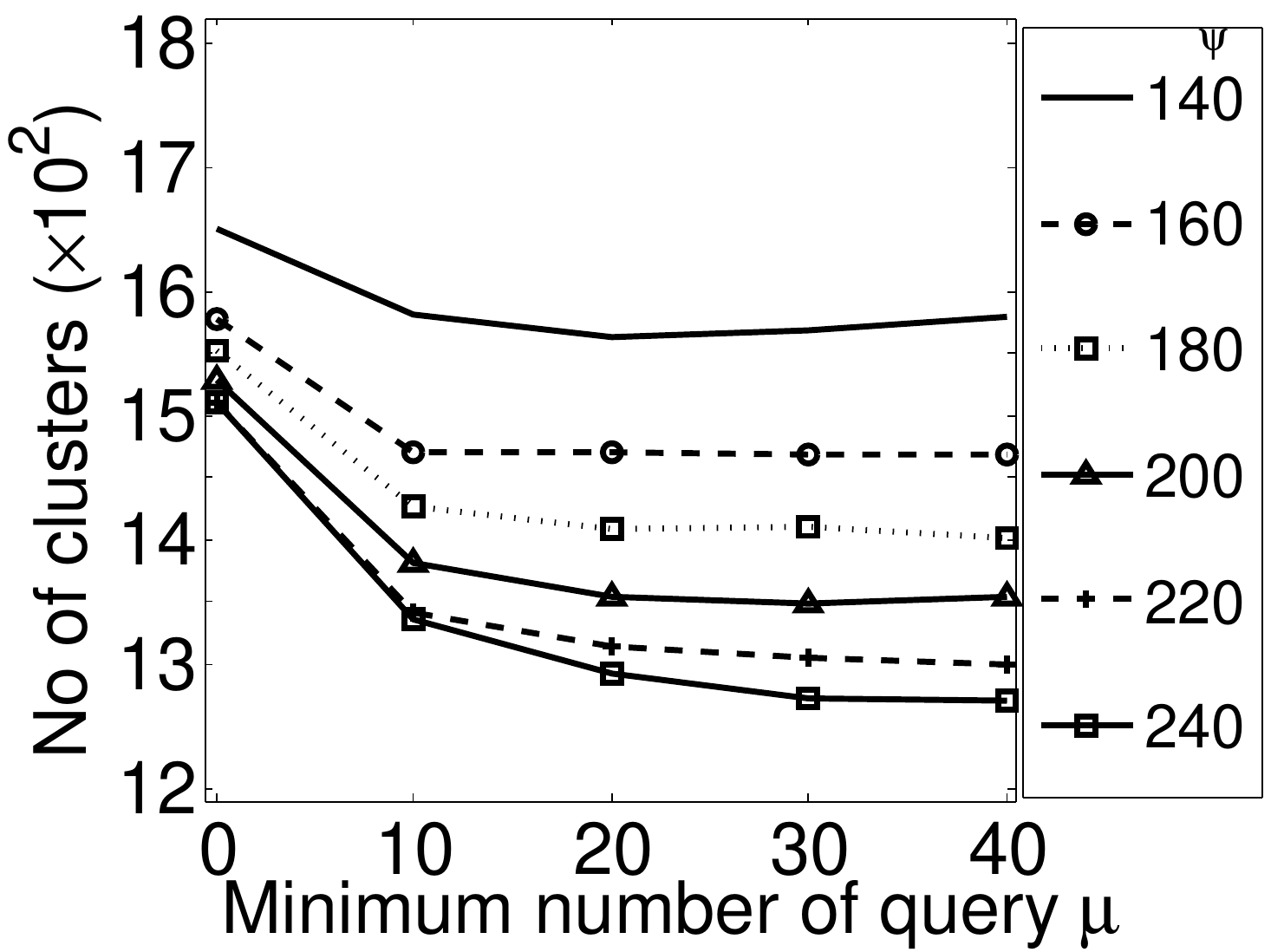}
\caption{Effect of $\mu$, for $\Delta$ = 80.} \label{fig:minq:vs:c:n80}
\end{figure}

Figure~\ref{fig:minq:vs:c:n80} shows the effect of minimum number of queries required to form a new cluster for $\Delta = 80$. We see that when
$\mu=0$ the number of clusters is very high. It decreases with the increase of the value of $\mu$ and remains almost constant when we choose values
closer to half the value of generated queries in each clusters. For experiment we generate clusters with 50 queries each. Thus we can see that for
both $\mu=20$ and $\mu=30$ the difference of number of clusters is insignificant. If we further increase the value of $\mu$ to 40, the time required
for clustering is increased. In this case most of the queries are not classified in the first phase, so more time is required to classify. Also a
large number of clusters contain a very small number of queries created in the second phase of Algorithm~\ref{algo:Q-lineGrouping}
(Lines~\ref{algo:Q-lineGrouping}.17 -- \ref{algo:Q-lineGrouping}.26).

\vspace{-5mm}
\vspace{-5mm}

We have shown that the number of unclustered queries for different values of $\mu$. Around 20\% of queries remain unclustered
when $\mu = 40$. For $\mu$ values of 20 and 30, the number of unclustered Q-lines are approximately 10\%. Again, Figure~\ref{fig:psi:vs:c} shows that
for $\mu = 30$, the effect of $\Delta$ and $\psi$ is minimum. When $\psi$ equals to 160, we have a less number of cluster count. Consider all these
performance issues we choose $\Delta = 80$, $\psi = 160$ and $\mu = 30$ as default values for our performance evaluation experiments.

\begin{figure}[htbp]
  \begin{center}
    \begin{tabular}{cccc}
         \hspace{-2mm}
      \resizebox{31mm}{!}{\includegraphics{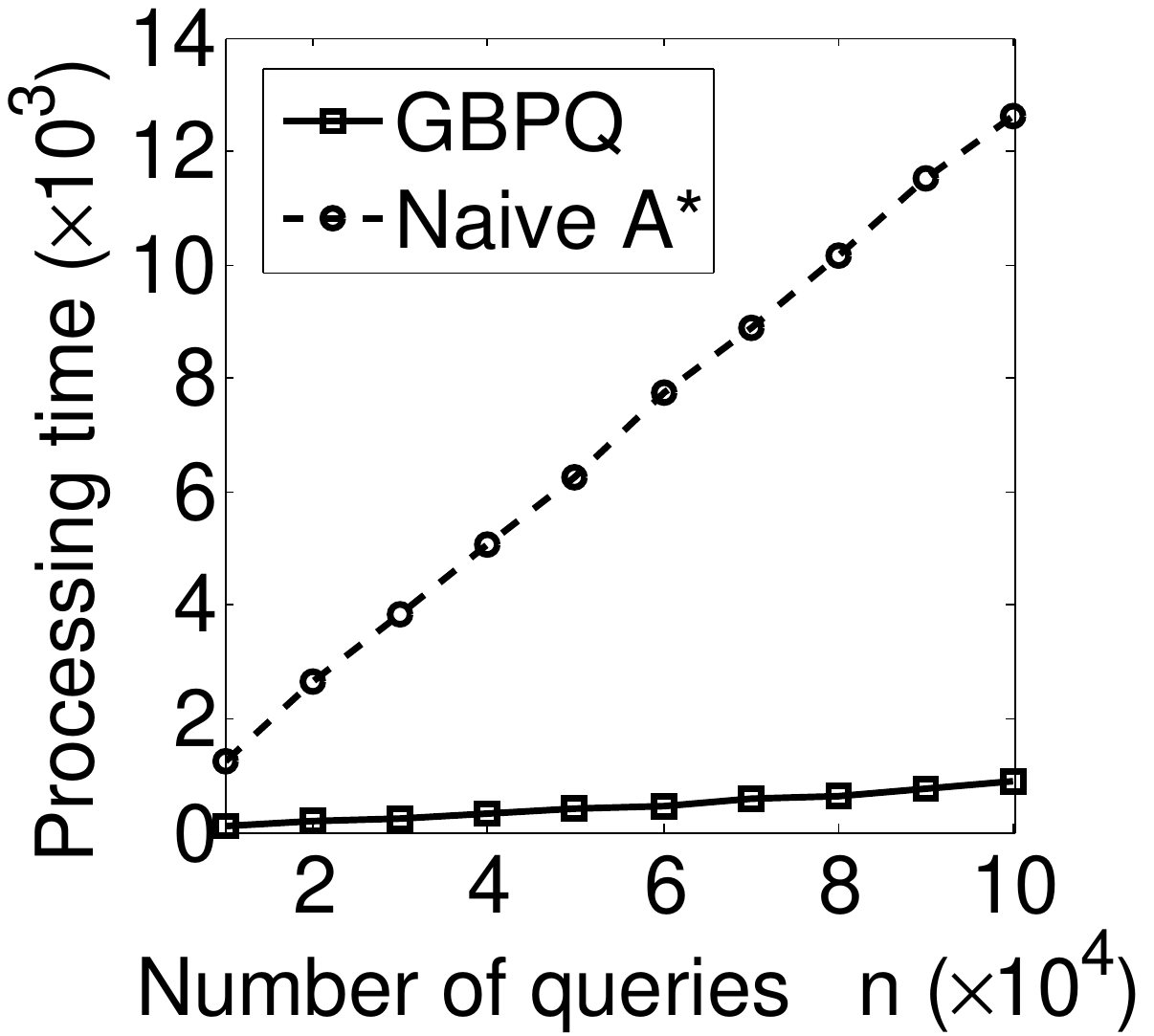}} &
         \hspace{-2mm}
      \resizebox{30mm}{!}{\includegraphics{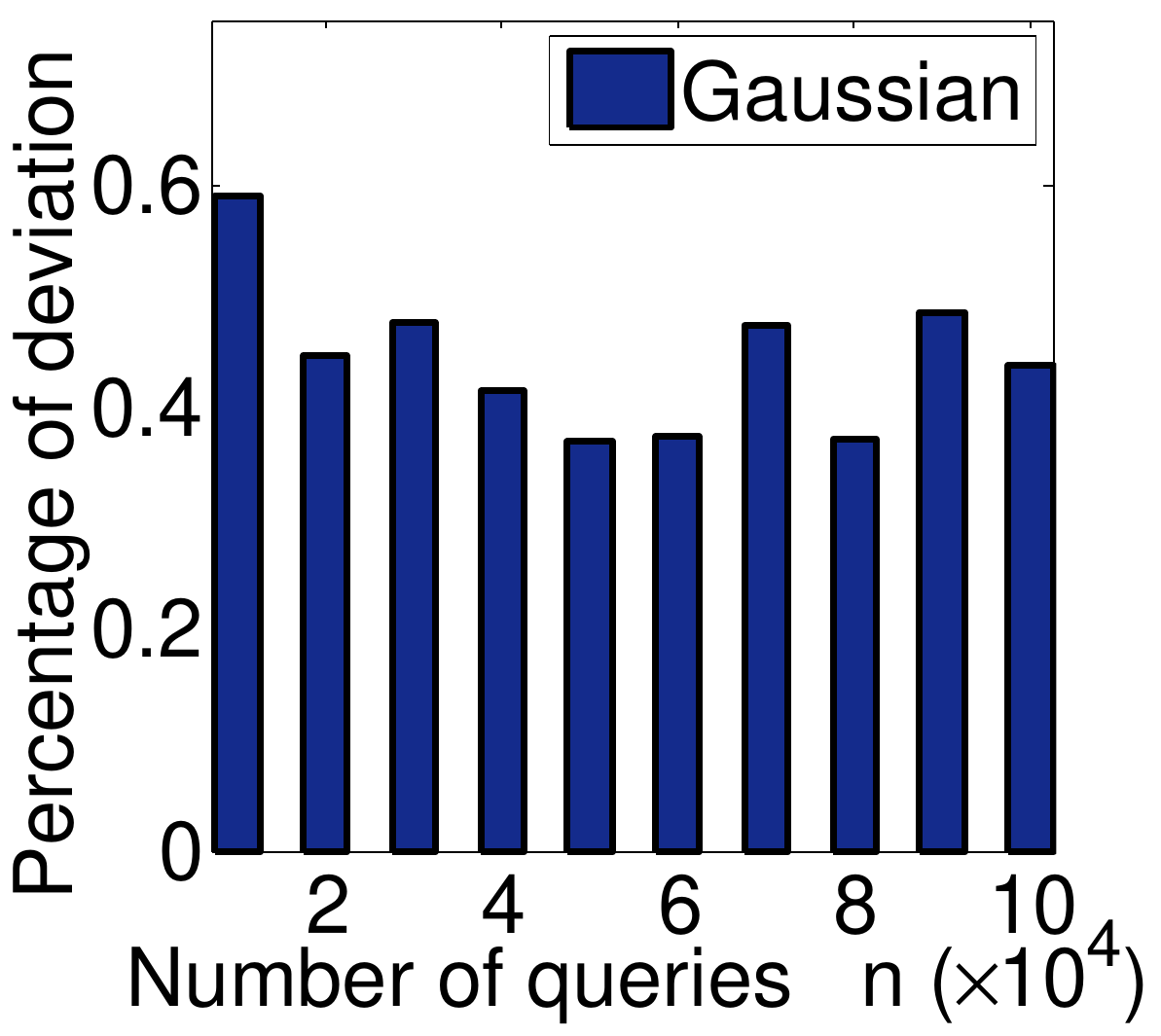}} & 
         \hspace{-2mm}
        \resizebox{31mm}{!}{\includegraphics{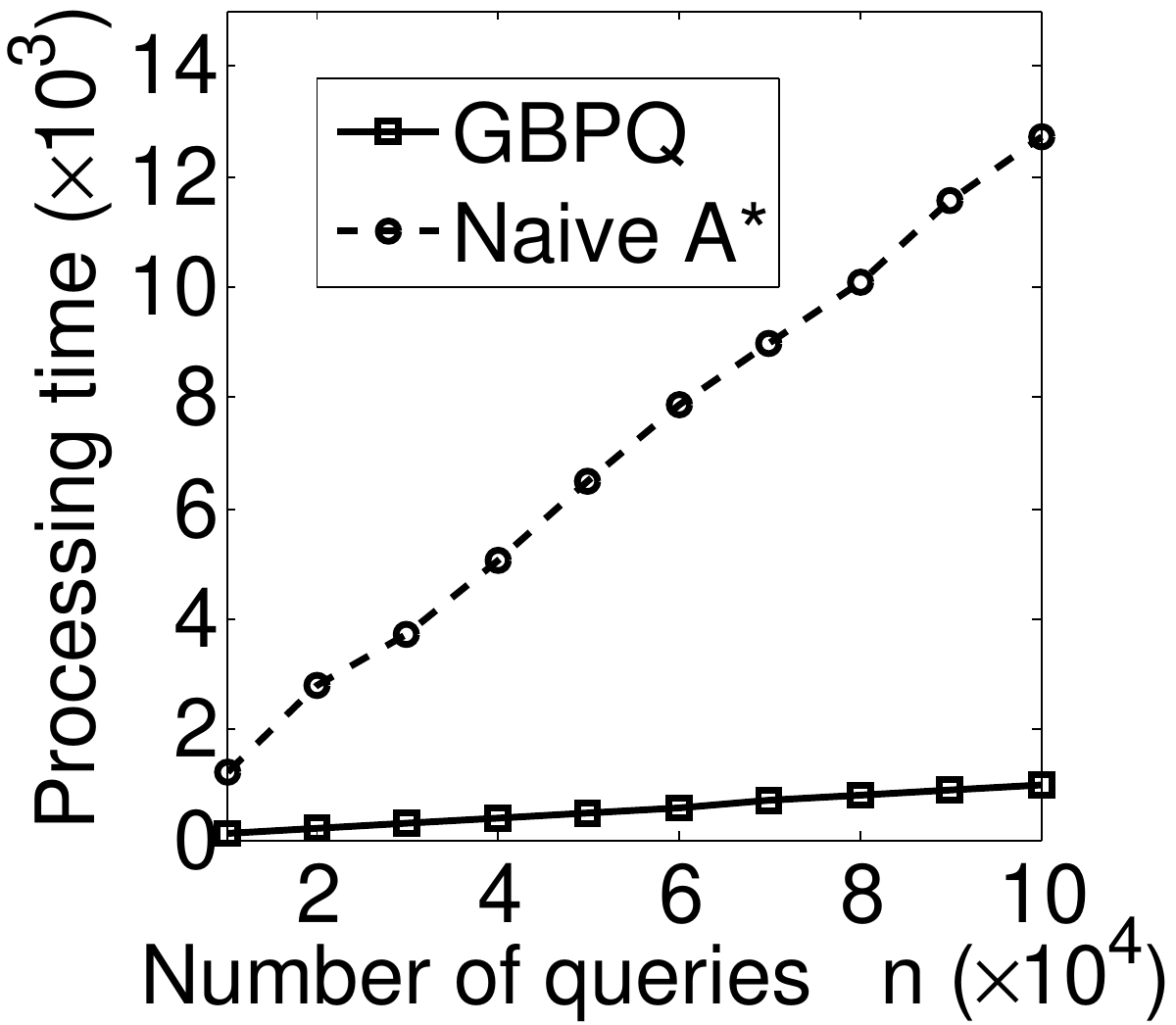}} &
         \hspace{-2mm}
      \resizebox{30mm}{!}{\includegraphics{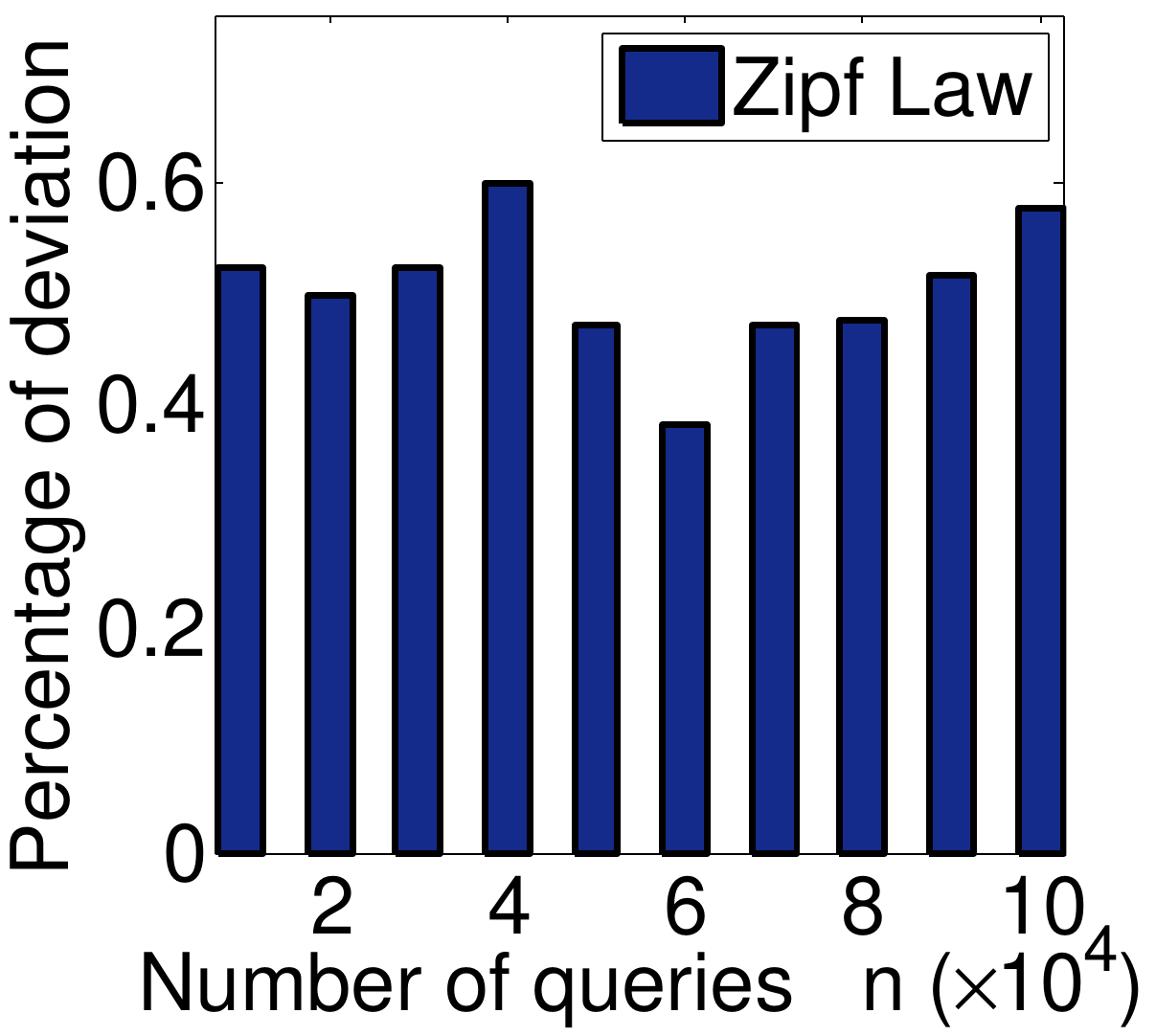}} \\
      \scriptsize{(a)\hspace{0mm}} & \scriptsize{(b)} & \scriptsize{(c)\hspace{0mm}} & \scriptsize{(d)}
        \end{tabular}
        \vspace{-4mm}
   \caption{ Effect of number of queries (a-b) Gaussian and (c-d) Zipf distributions.}
\label{fig:q:vs:td:nz}
  \end{center}
   \vspace{-10mm}
\end{figure}


\subsection{Performance Evaluation}
In this section, we show the performance of our algorithm based on the two performance metrics: processing time and the average percentage of deviation of the answer from
actual shortest path. Processing time is the total query response time including the clustering time for a given number of queries. Clustering time is about 6 seconds on average for 50000 queries and shows a linear relationship with number of queries for the selected parameter values. Average deviation percentage is calculated with the formula $\frac{d_tr - d_ta}{d_ta}\times 100\%$ where $d_tr$ is the total distance returned and $d_ta$ is the actual total shortest path distance for given number of queries.
 We vary the number of queries $n$, the minimum query distance, i.e., the distance between the source point and destination
point of each query, and the window size $\omega$, while comparing our approach with the naive approach. We determine the minimum query distance of a
query by multiplying $d_c$ with  $100\sqrt{\omega}$. We calculate the minimum query distance as a function of $\omega$, because it allows us to have
queries within the same window when the window size is larger and span the queries into several windows when window size is smaller. The range, step
size and default values for these parameters are listed in the Table~\ref{table:criteria}.

\subsubsection{Effect of number of queries:}
We vary the number of queries $n$ in the range of 10000 to 100000 with a step size of 10000 units. For both Gaussian and Zipf distributions we see
that the processing times for GBPQ rises slightly with the increase of the values of $n$ (Figure~\ref{fig:q:vs:td:nz}a,~\ref{fig:q:vs:td:nz}c).
Whereas, for the naive approach, the processing time increases significantly with the increase in the number of queries. When the value of $n$ is
10000, the processing times for GBPQ  and the naive approach are approximately 100s and 1000s, respectively. With an increased value 100000 of $n$,
the processing times for GBPQ  and the naive approach are approximately 1000s and 13000s, respectively. Thus, we see that our algorithm outperforms
the naive approach by a greater margin for an increased value of $n$. On an average GBPQ is twelve times faster than the naive approach.  Moreover,
our experimental results show that on an average the deviation of the answer path returned by GBPQ from the actual shortest path is only around $0.4\%$ in
case of Gaussian distribution (Figure~\ref{fig:q:vs:td:nz}b) and around $0.5\%$ in case of Zipf (Figure~\ref{fig:q:vs:td:nz}d).

\subsubsection{Effect of query distance:}
\begin{figure}[htbp]
  \begin{center}
    \begin{tabular}{cccc}
         \hspace{-5mm}
      \resizebox{30mm}{!}{\includegraphics{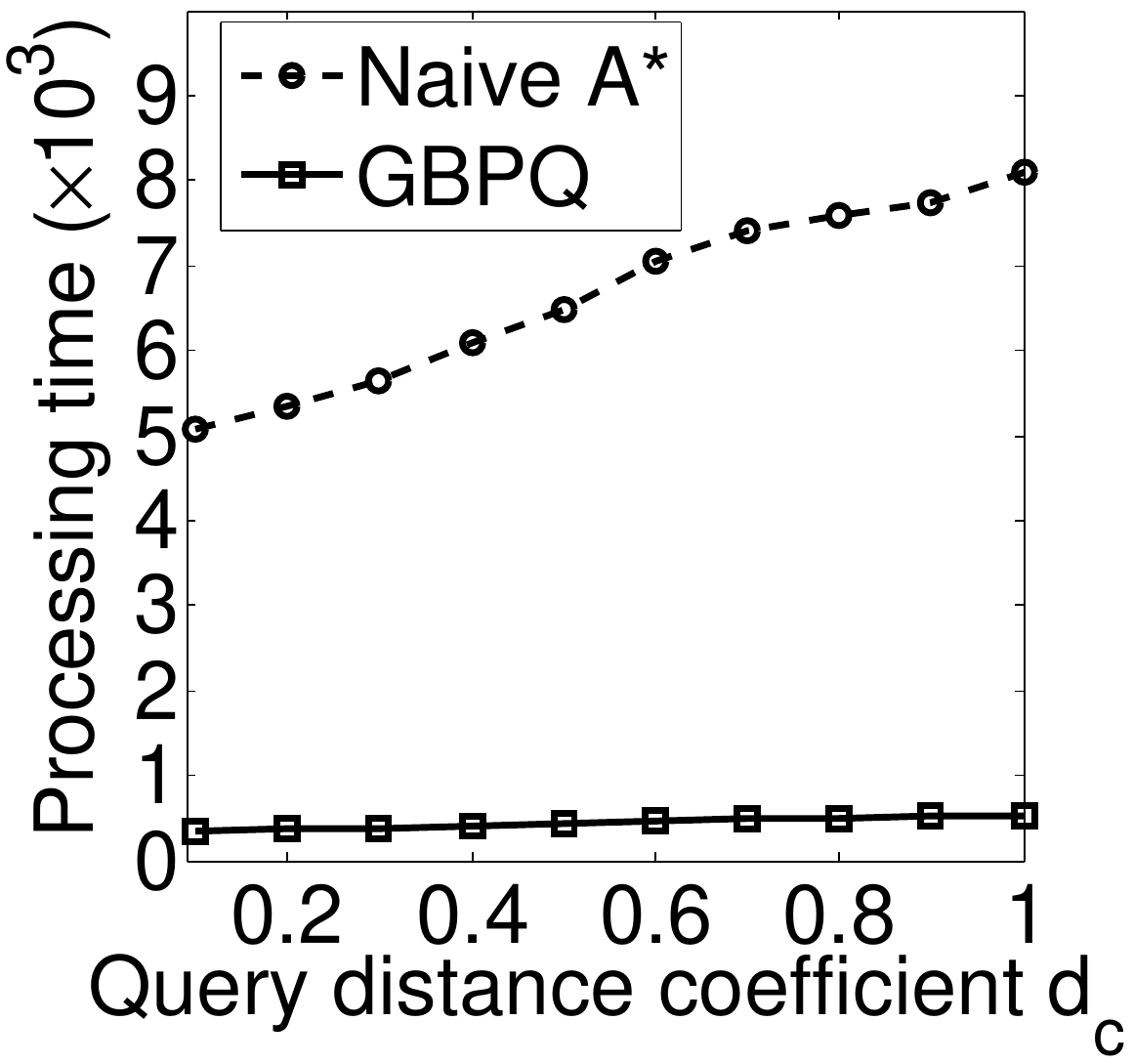}} &
         \hspace{-2mm}
      \resizebox{30mm}{!}{\includegraphics{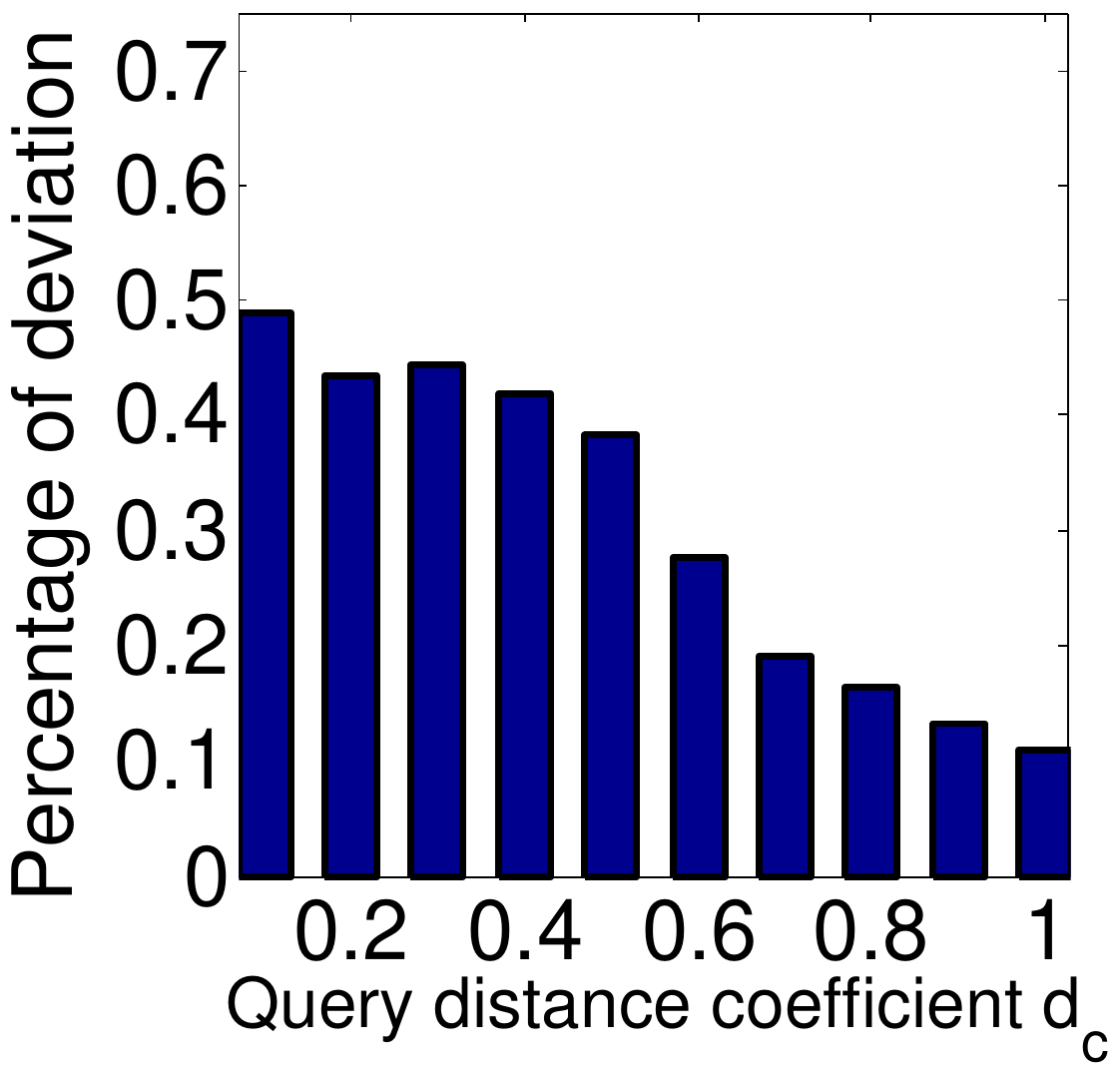}} &

         \hspace{-2mm}
      \resizebox{30mm}{!}{\includegraphics{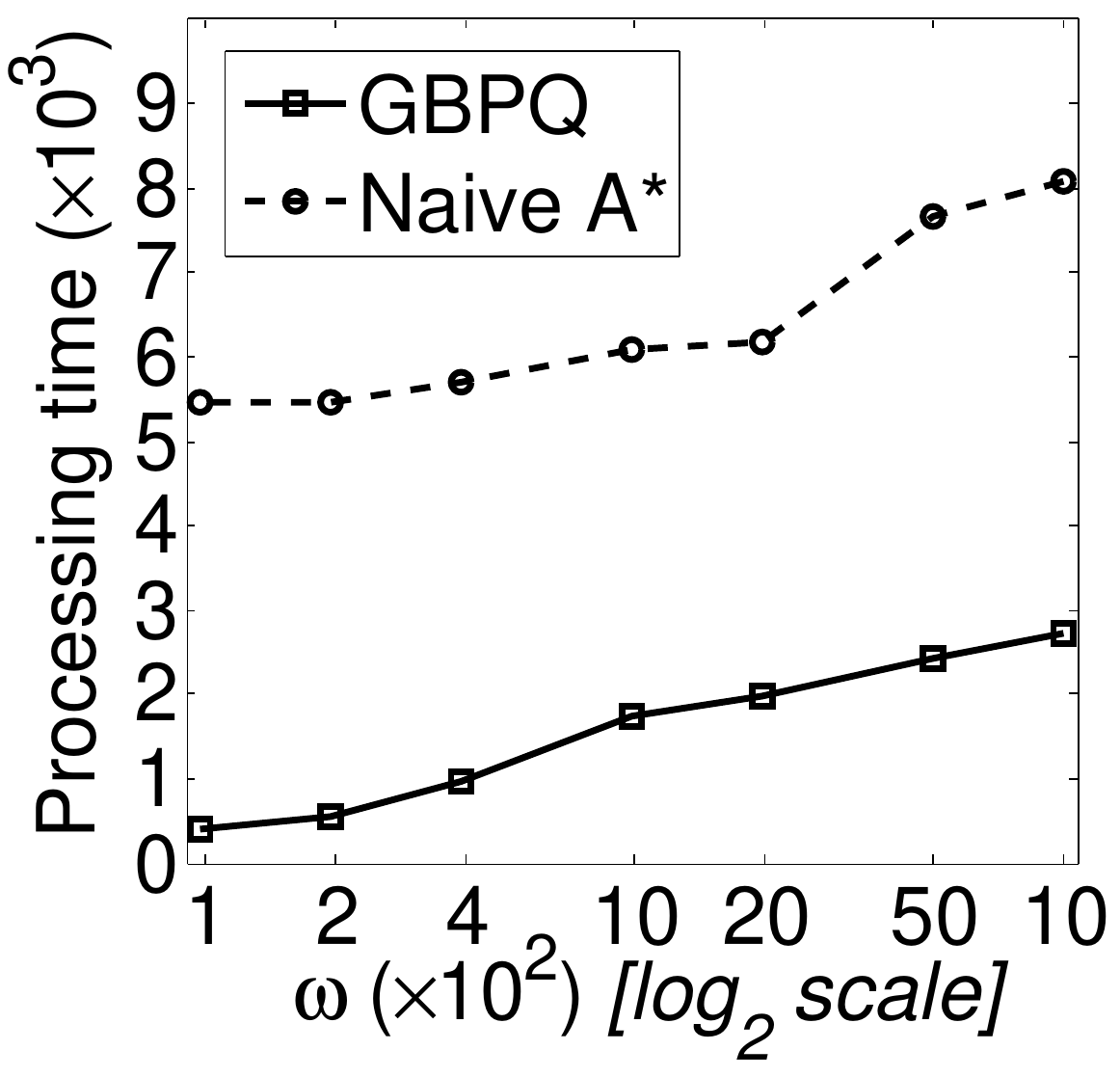}} &
         \hspace{-2mm}
      \resizebox{30mm}{!}{\includegraphics{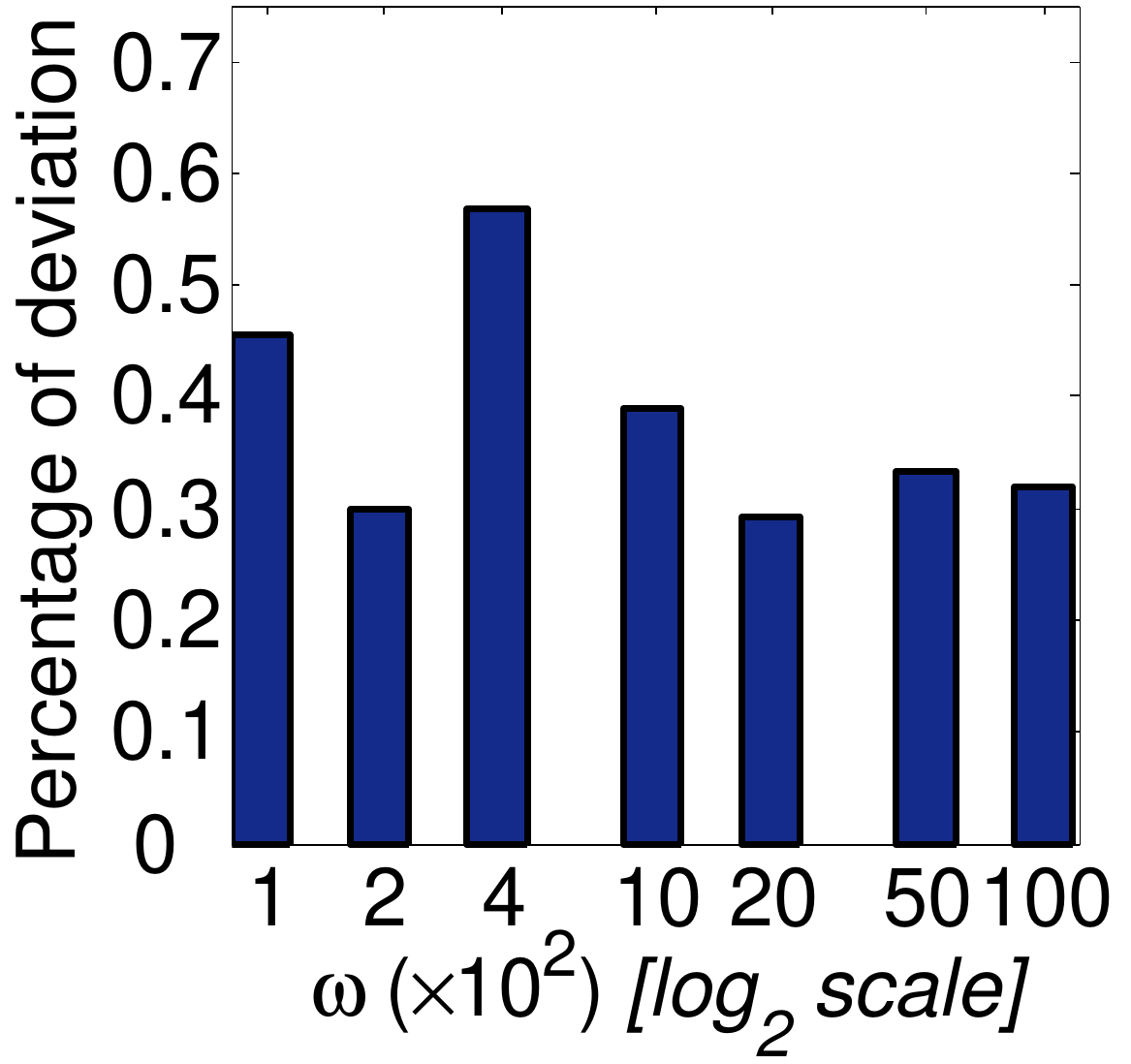}} \\
      \scriptsize{(a)\hspace{0mm}} & \scriptsize{(b)} & \scriptsize{(a)\hspace{0mm}} & \scriptsize{(b)}
        \end{tabular}
   \caption{Effect of the minimum query distance, (a) processing time (b) percentage of deviation; Effect of window size, (c) processing time (d) percentage of deviation}
\label{fig:mw:vs:td}
  \end{center}
   \vspace{-5mm}
\end{figure}

In this set of experiments, we compare our approach with the naive approach by varying the minimum query distance. For this, we vary the minimum
query distance coefficient $d_c$ for generating queries in our experiments and measure the processing time for GBPQ and the naive approach.
Figure~\ref{fig:mw:vs:td}a shows the results for Gaussian distribution of query points. We see from the figure that for GBPQ the processing time
slowly increases with the increase of the value of $d_c$. On the other hand, the processing time increases significantly for the naive approach as
the value of $d_c$ increases. This is because, a higher value of $d_c$ corresponds to a longer distance between the source and destination and the
number of nodes traversed for such a query is higher than that of the query who has a smaller $d_c$. Thus, the processing time increases with the
increase of the value of $d_c$. The accuracy of our GBPQ increases with the increase of $d_c$. The percentage of deviation of the answer path
from the actual path reduces from $0.5\%$ to  $0.1\%$ when the query distance increases from 1000 units to 10000 units (Figure~\ref{fig:mw:vs:td}b).

The results for Zipf query distribution (not shown) shows exactly similar behavior as Gaussian distribution.

\vspace{-3mm}

\vspace{-10mm}
\subsubsection{Effect of window size:}
In this set of experiments, we vary the window length $\omega$ and compare the performance of our approach with the naive approach.
Figure~\ref{fig:mw:vs:td}c shows the processing time of GBPQ and the naive approach for Gaussian distribution of query points. We see the processing
time increases with the increase of the value of $\omega$. For example, when there is no partition i.e. the window length is equal to the length of
the data space, the processing time of GBPQ is 62.3\% lower compared to that of the naive approach, whereas, for 100x100 sized windows the processing
time of GBPQ is 92\% lower compared to that of the naive approach.


Figure~\ref{fig:mw:vs:td}d shows the deviations of GBPQ's answers for different values of $\omega$. We find that the deviation is restricted
between 0.29\% and 0.56\%. When the value of $\omega$ is low, the number of clusters created is high. This causes a lower rate of deviation for a
higher value of $\omega$. For a higher value of $\omega$,  we can see a slightly higher deviation than that of a lower $\omega$ as the number of
clusters created is low in this case.

The results for Zipf query distribution (not shown) shows exactly similar behavior as Gaussian distribution.


\section{Conclusion}

In this paper, we have proposed a group based approach for processing a large number of simultaneous path queries in a road network. Our approach is based on a novel clustering technique that groups queries based on the similarities of their {\itshape Q-lines}. We introduce two concepts: the distance function and the areas of influence, that helps us to effectively cluster similar queries and execute them as a group. Our group based approach to evaluate a large number of simultaneous path queries provides a cost-effective solution with reduced computational overhead.

Extensive experimental studies show the efficiency and validate the effectiveness of our proposed algorithm. The group based heuristic in our approach reduces the computational overhead significantly and thereby answers a large number of simultaneous queries in real time. Our experiments have shown that on an average our shared execution approach is ten times faster than a traditional approach, where each path query is evaluated individually. Our approach achieves this huge superiority at the cost of sacrificing the accuracy by a negligible amount of 0.5\% in the worst case.

\bibliographystyle{abbrv}
\bibliography{references}

\end{document}

\section{Acknowledgments}
This section is optional; it is a location for you
to acknowledge grants, funding, editing assistance and
what have you.  In the present case, for example, the
authors would like to thank Gerald Murray of ACM for
his help in codifying this \textit{Author's Guide}
and the \textbf{.cls} and \textbf{.tex} files that it describes.
\end{document}